\documentclass[twocolumn,showpacs,amsmath,amssymb,prb]{revtex4}

\usepackage{amsmath}
\usepackage{amssymb}
\usepackage{graphicx}
\usepackage{dcolumn}
\usepackage{bm}

\usepackage{pstricks,pst-node,pst-text,pst-3d}
\usepackage{subfigure}
\def\be{\begin{eqnarray}}
\def\ee{\end{eqnarray}}
\def\ba{\begin{array}}
\def\ea{\end{array}}
\def\nn{\nonumber}
\begin{document}

\title{The visibility of IQHE at sharp edges:\\ Experimental proposals based on interactions and edge electrostatics}
\author{U. Erkarslan$^1$}
\author{G. Oylumluoglu$^1$}
\author{M. Grayson$^2$}
\author{A. Siddiki$^{3,4}$}
\address{$^1$ Department of Physics, Faculty of Arts and
Sciences, Mugla University, 48170-Kotekli, Mugla, Turkey}
\address{$^{2}$ Electrical Engineering and Computer Science, Northwestern University, Evanston, IL 60208 USA}
\address{$^{3}$ Department of Physics, Faculty of Sciences, Istanbul University, 34134 Vezneciler, Istanbul, Turkey}
\address{$^{4}$ Department of Physics, Faculty of Arts and Sciences, Harvard University, Cambridge, 02180 MA USA}
\begin{abstract}
The influence of the incompressible strips on
the integer quantized Hall effect (IQHE) is investigated, considering a cleaved-edge overgrown (CEO) sample as an experimentally realizable sharp edge system. We propose a set of experiments to clarify the distinction between the large-sample limit when bulk disorder defines the IQHE plateau width and the small-sample limit smaller than the disorder correlation length, when self-consistent edge electrostatics define the IQHE plateau width. The large-sample or bulk QH regime is described by the usual localization picture, whereas the small-sample or edge regime is discussed within the compressible/incompressible strips picture, known as the screening theory of QH edges. Utilizing the unusually sharp edge profiles of the CEO samples, a Hall bar design is proposed to manipulate the edge potential profile from smooth to extremely sharp. By making use of a side-gate perpendicular to the two dimensional electron system, it is shown that the plateau widths can be changed or even eliminated altogether. Hence, the visibility of IQHE is strongly influenced when adjusting the edge potential profile and/or changing the dc current direction under high currents in the non-linear transport regime. As a second investigation, we consider two different types of ohmic contacts, namely highly transmitting (ideal) and highly reflecting (non-ideal) contacts. We show that if the injection contacts are non-ideal, however still ohmic, it is possible to measure directly the non-quantized transport taking place at the bulk of the CEO samples. The results of the experiments we propose will clarify the influence of the edge potential profile and the quality of the contacts, under quantized Hall conditions.
\end{abstract}


\maketitle
The integer quantized
Hall effect (IQHE) is observed in a two-dimensional electron gas (2DEG) subjected to a strong perpendicular magnetic field $B$, with signatures in the longitudinal $R_L$ and transverse $R_H$ resistances.~\cite{vKlitzing80:494}  At certain magnetic field intervals $R_L$ vanishes and $R_H$ is quantized. Surprisingly, the theories that elucidate the IQHE are still under discussion, even today.~\cite{Halperin:10} Two main schools emerged in explaining the IQHE, namely
the bulk~\cite{Laughlin81,Kramer03:172} and the
edge~\cite{Halperin82:2185,Buettiker86:1761,Chklovskii92:4026} pictures. They are
thought to contrast each other in describing the current distribution. The former assumes that the transport is at the bulk and quantization is determined by localization effects, whereas the latter neglects the effects of disorder and assumes 1D ballistic channels at the edge, carrying the quantized current. Though early arguments took a contrasting view,~\cite{Chklovskii92:4026} it is currently widely accepted that edge currents flow losslessly within the incompressible strips.~\cite{Chang90:871,siddiki2004}

The compressible/incompressible strips result electrostatically from a quantizing magnetic field in combination with self-consistent direct Coulomb interactions under the constraint of electrostatic boundary conditions. The magnetic field quantizes the density of states into highly degenerate Landau levels separated from each other by cyclotron or Zeeman energy gaps.  If the
Fermi energy is pinned to one of these Landau levels for a finite width parallel to the edge of the sample, strip is called compressible, due to the high degeneracy at the density of states (DOS). If the Fermi energy resides {\em between} Landau levels, the strip is called incompressible, since there are no available states at the Fermi energy for the entire width of this strip. It is standard to define a dimensionless parameter called the filling factor $\nu$, which gives the ratio of the electron density $n_{\rm el}$ to the density of magnetic flux quanta $n_{\Phi}$. One can express the filling factor as $\nu=n_{\rm el} / n_{\Phi} = 2\pi \ell^{2}_{B}n_{\rm el}$, where $\ell_{B}=\sqrt{\hbar/eB}$ is the magnetic length. An integer $\nu$ implies that, all the Landau
levels below the Fermi energy are fully occupied, hence the system is in an incompressible state. Otherwise the Landau level is pinned to the Fermi energy and is partially occupied, therefore the system is compressible. The properties of these strips are investigated intensely in the literature, both theoretically~\cite{Chklovskii92:4026,Halperin94:etchedge,Oh97:13519,Sefa08:prb,Lier94:7757,Akera06:,Guven03:115327,SiddikiEPL:09} and experimentally.~\cite{Wei98:1674,Ahlswede01:562,Ahlswede02:165,Grayson05:016805,Matt:ceo1,Dahlem10:contact,deviatov:10,Sailer:10,afif:njp2} The theoretical investigations, known as the screening theory, aim to clarify the effects of the boundary electrostatics,~\cite{Chklovskii92:4026,Halperin94:etchedge,Oh97:13519,Sefa08:prb} the temperature~\cite{Lier94:7757,Akera06:} and the current~\cite{Guven03:115327,SiddikiEPL:09} on the formation of the incompressible strips. More recently, the behavior of the incompressible strips near the contacts has been investigated experimentally~\cite{Dahlem10:contact} and theoretically.~\cite{Deniz10:contact} These investigations helped to resolve the long standing question of what characteristics define an ideal contact. However, the effect of electrostatic boundary conditions on the formation of incompressible strips far from contacts is left unresolved.

The objective of this work is two-fold: first, we explore the existence of incompressible strips in the presence of infinitely sharp confinement walls. Our discussion is based on the existing experimental literature~\cite{Ahlswede02:165,Grayson05:016805,Dahlem10:contact,afif:njp2} and semi-classical calculations.~\cite{siddiki2004,SiddikiEPL:09} We show that, these strips vanish at sharp edges due to overlap of the quantum mechanical wavefunctions across the incompressible strip remnant. Exploiting the findings of the interaction theory of the IQHE~\cite{siddiki2004} and local probe experiments,~\cite{Ahlswede02:165} we claim that the longitudinal resistance evolves differently on sharp and smooth edges when transitioning between QHE plateaus. Second, we show that the Hall resistance is non-quantized even at the plateau regime due to the evanescent bulk current. This behavior depends on the contact quality. Our work is organized as follows: In the following Section, we briefly introduce the material and geometrical properties of the sample proposed here. The first part of Section II is spared for the formulation of the screening theory, where we also present results of self-consistent numerical calculations. Second part highlights the essentials of local probe and CEO experiments, followed by a brief discussion of the transport model, Sec.\ref{sec:ohmslaw}. The effect of ideal/non-ideal contacts on transport, considering a varying magnetic field is investigated in Sec.\ref{sec:contacts}. There we show that, in the case of ideal contacts one expects to observe differences between generic and CEO samples, when measuring either the local electrochemical or the longitudinal resistances. For the non-ideal contacts we predict that, some of the current is scattered to bulk due to the absence of incompressible strip at the CEO. Hence, the Hall resistance is not quantized even at the plateau regime. We discuss the effects of the sweeping direction and the orientation of magnetic field on transport, which induces a hysteresis in Sec.\ref{sec:other}.

\begin{figure}[t]
\includegraphics[scale=0.20]{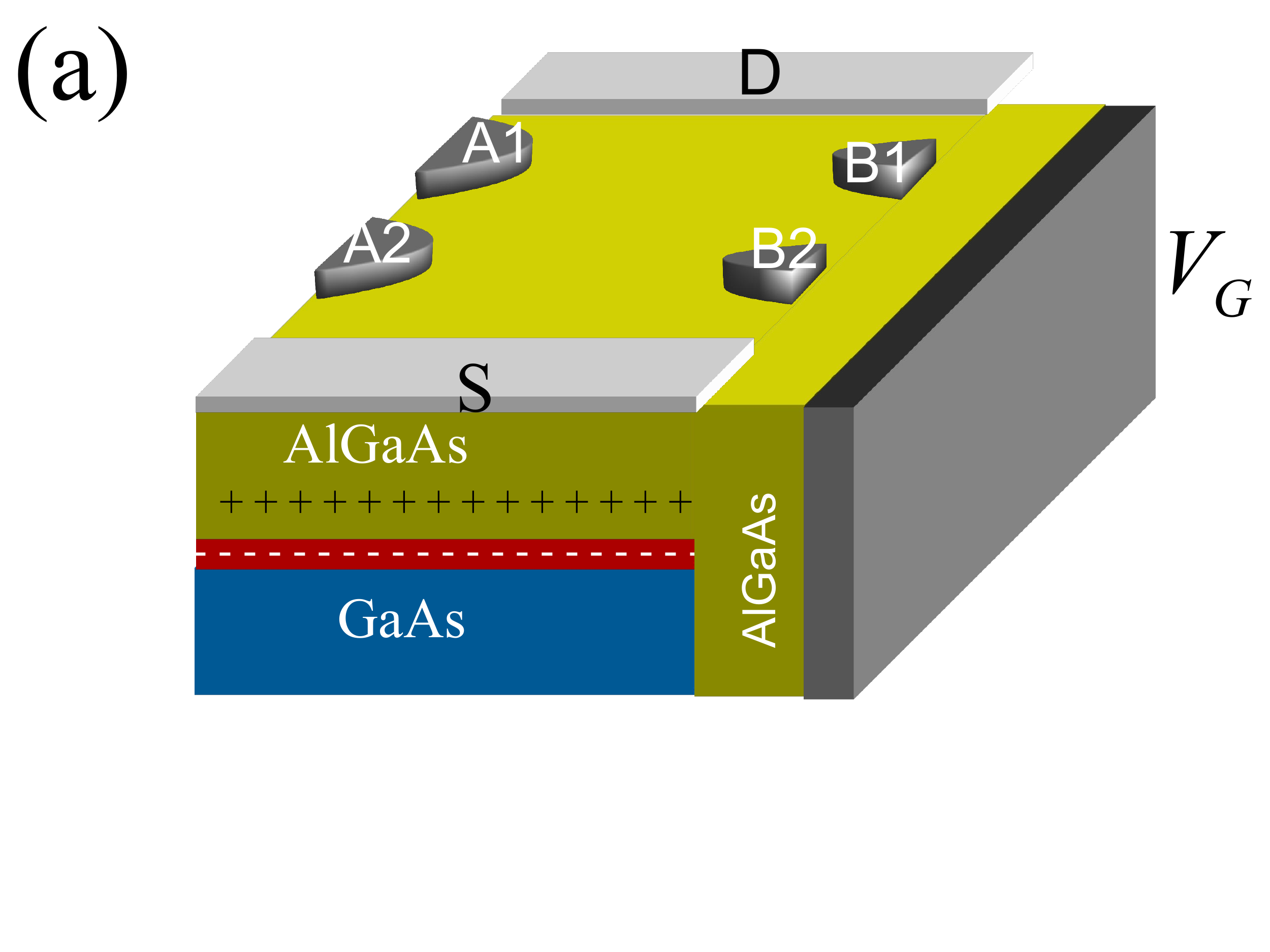}
\includegraphics[scale=0.20]{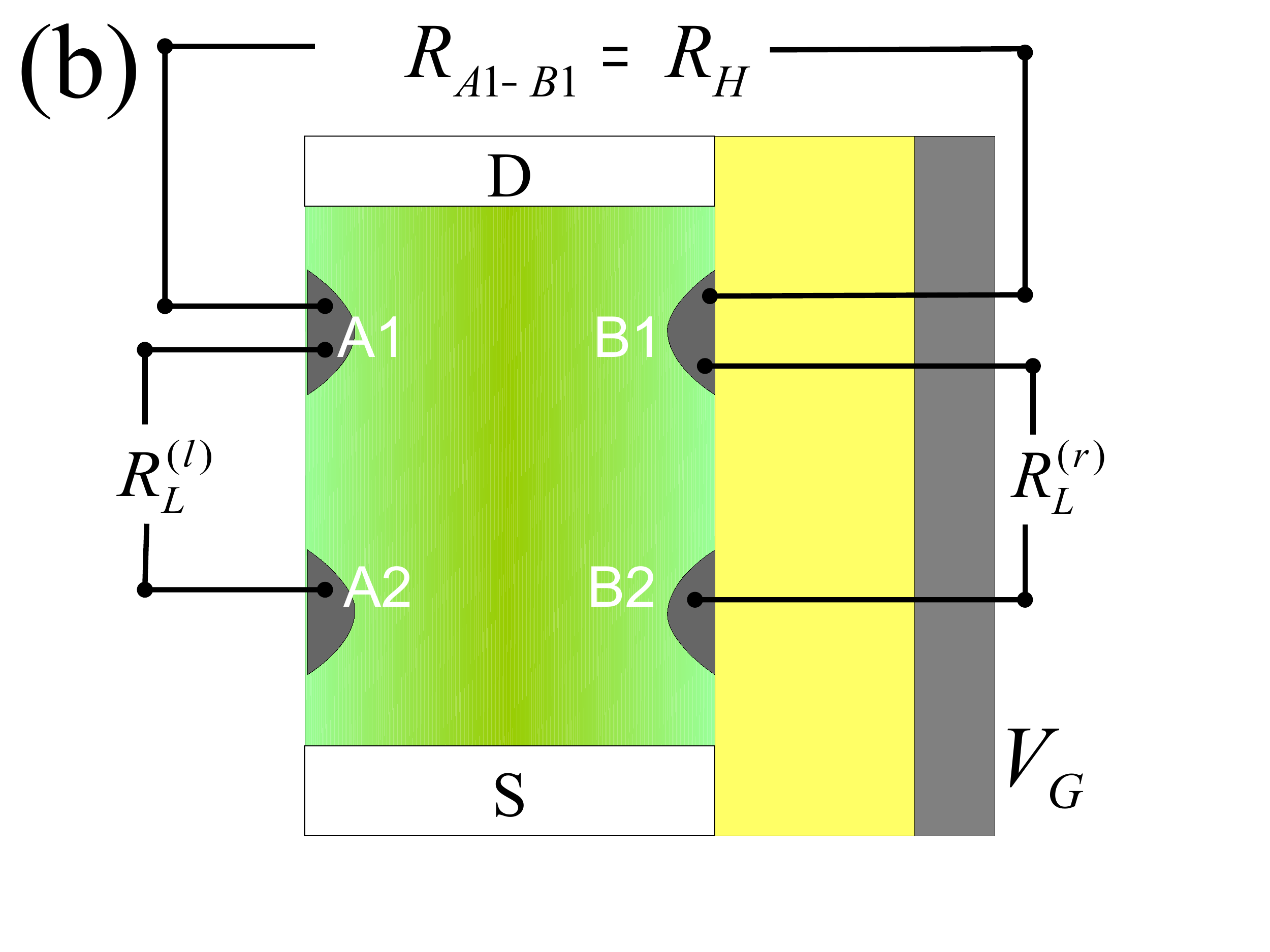}
\caption{\label{fig:fig1}The side-view (a) and top-view (b) projections of the sample design. Hall bar is defined by etching on the left hand side (LHS), generating a relatively smooth potential profile, whereas on the right hand side (RHS) the CEO edge serves as a steep potential. The steepness is manipulated by the
side gate (gray region), by applying a negative potential. The illustration is not to scale.}
\end{figure}
\section{The wafer and the sample geometry}
In standard molecular beam epitaxy (MBE), the crystal is
grown in the $z$ direction layer by layer. However, well-established experimental techniques are able to grow materials in different crystal
directions in sequential steps. As a first step a usual MBE growth
process is performed then the crystal is removed from the chamber,
thinned, scribed, returned to the chamber mounted at a 90 degree
angle, then cleaved in-situ and regrown. This process is known as
the cleaved edge overgrowth (CEO) technique.~\cite{CEO:pfeifer} The CEO samples are used to investigate the edge
properties of the quantum Hall systems via momentum-resolved tunneling
experiments, utilizing the second 2DEG residing perpendicular to the Hall system.~\cite{Grayson05:016805,Matt:ceo1} These
tunneling experiments show that the CEO edge provides an extreme sharp potential approximating an infinite wall.  The sample structure proposed here does not involve tunneling,  but rather probes the sharpness of the QHE edges via transport alone.

A schematic presentation of the crystal is depicted in Fig.~\ref{fig:fig1}. The Hall bar lies in the $xy$ plane, obtained by usual MBE growth, whereas on the right hand side of the crystal an additional AlGaAs layer is grown. This is the CEO edge and is capped by a metallic gate. The gate covers all the surface and is kept at a fixed potential $V_G$. To measure the global resistances and to drive an external current $I$ through the 2DEG, ohmic contacts are deposited on the top surface of the system. The voltage probe contacts are labeled by A1, A2, B1 and B2 to measure $R_H$ and $R_L$, whereas the current contacts S and D denote the source and the drain. In our discussion of the visibility of the IQHE, we will describe transport in such a CEO sample. We consider a system, where at one edge (CEO edge) a side gate resides, meanwhile the opposing edge is defined by standard etching process.
\section{The formulation of the screening theory, relevant experiments and the complementary transport model}
This section aims to provide necessary theoretical and experimental background, to discuss the transport properties of the proposed CEO samples. Our theory discussion is composed of two parts: first, we re-introduce the analytical formulation behind the self-consistent screening calculations and discuss the existence of incompressible strips. The second part presents the numerical results. Next, we summarize the observations of two sets of experiments relevant for our discussion. The distinction between the small-sample edge IQHE and the large-sample bulk IQHE, is clarified after we discuss the results of local probe experiments and corresponding theory results. In the last part, we briefly recount the complementary transport calculations utilized by the screening theory to elucidate IQHE.
\subsection{The calculation scheme}
Here, we briefly re-introduce the calculation scheme to obtain electrostatic potential and electron density distributions of a 2DEG. Let us assume that, the electrons are confined at the $xy$ plane due to an external potential $V_{\rm conf}(x,y)$, which results from remote donors and metallic gates. At zero temperature and in the absence of a perpendicular magnetic field, this confinement potential will be filled with electrons up to the Fermi energy $E_F$ continuously by the virtue of constant density of states $D_0~(=m/\pi \hbar^2)$, specific to 2D. Taking into account the mutual (direct) Coulomb interaction within a mean-field approximation, one obtains the Hartree potential (energy) as
\be V_{\rm Ha}(x,y)=\frac{e^2}{\kappa}\int_An_{\rm el}(x,y)K(x,y;x',y')dxdy,\ee
where $\kappa$ is the dielectric constant of the material, $A$ is the area where the 2DEG is confined and $K(x,y;x',y')$ is the solution of the Poisson equation for the given boundary conditions, to be specified later. As a direct consequence, the total electrostatic potential energy that an electron experiences is
\be V_{\rm T}(x,y)=V_{\rm conf}(x,y)+V_{\rm Ha}(x,y).\ee In the next step, one should calculate the new electron distribution regarding this potential via solving the single particle Schr\"odinger equation. For the moment, let us assume a generic Hamiltonian $H$ and depict the energy eigen functions by $\phi_{\alpha}(x,y)$ and eigen values by $E_{\alpha}$. The effect of temperature on electron occupation can be incorporated utilizing the Fermi-Dirac function $f(E_{\alpha}, E_F,T)$, then the electron density distribution is described as
\be n_{\rm el}(x,y)=\sum_{\alpha}|\phi_{\alpha}(x,y)|^2 f(E_{\alpha},E_F,T).\label{eq:edensity}\ee

Let us specify $H$ to be the single particle Hamiltonian of a spinless electron subject to perpendicular magnetic field, given by
\be H=H_B=\frac{1}{2m}(\textbf{p}-\frac{e}{c}\textbf{A}(x,y))^2+V_{\rm T}(x,y), \ee
where $\textbf{p}$ is the momentum and $\textbf{A}(x,y)$ is the vector potential generating the magnetic field. Here, we assume a translational invariance and the vector potential is expressed in the Landau gauge. If we neglect the potential term for the moment, then the solution of the Schr\"odinger equation gives the energy eigen values to be $E_{n,X}=\hbar\omega_c(n+1/2)$, known as the Landau levels, and the wavefunctions as, $\phi_{n,X}(x,y)=N\exp(iky)\exp(-(x-X)^2/2\ell_B^2)H_n((x-X)/\ell_B)$. Here, $n$ is the Landau level index, $N$ is a normalization factor, $k$ is the quasi-continuous momentum in $y$, $X~(=-\ell_B^2k)$ specifies the orbit-center coordinate and $\omega_c=eB/m$ is the cyclotron frequency. Since, the eigen energies and values are known, one can obtain the electron density by Eq.~\ref{eq:edensity}.

Now one can calculate the new potential depending on the new electron distribution, via solving the Poisson equation. This procedure requires numerical self-consistency and is a formidable task in general. Despite this difficulty, such a numerical algorithm is available, which we also discuss in the following. However, we first re-introduce the non self-consistent analytical scheme to base our discussion on a simpler ground.
\subsubsection{The non self-consistent analytical approach}
To surpass the complications due to numerical self-consistency, Chklovskii et.al proposed an analytical calculation scheme to obtain potential and electron density distributions based on electrostatic arguments.~\cite{Chklovskii92:4026} There, the solution of the Poisson equation is obtained by the use of holomorphic functions considering in-plane metallic gates at the sides. The commonly used Thomas-Fermi approximation is invoked, which is adequate if the external potential varies smoothly within the magnetic length. In addition, the wavefunctions are replaced by Dirac-delta functions. It is assumed \emph{a priori} that, screening is perfect at the compressible strips and is poor at the incompressible strips. Hence, the external potential is completely screened at the compressible strips and total potential profile there is flat. At the compressible strips the density profile resembles that without $B$ field, given by
\be n_{\rm el}(x)=n_0\bigg(\frac{x-l_d}{x+l_d}\bigg)^{1/2},\ee
via the solution of the Poisson equation. Here, $n_0$ is the bulk electron density and $l_d$ is the depletion length determined by the potential at the in-plane gate.~\cite{Chklovskii92:4026} In contrast, at the dipolar incompressible strips the potential varies by an amount of cyclotron energy $\hbar\omega_c$, whereas the density is constant. Under these assumptions the width of the dipole incompressible strip is given by,~\cite{Chklovskii92:4026}
\be a_n=\sqrt{\frac{2\kappa\Delta E}{\pi^2e^2dn(x)/dx|_{x=x_n}}}, \ee
where $\Delta E$ is the single particle energy gap ($=\hbar\omega_c$, assuming spin degeneracy). In generic samples, the electron density distribution varies smoothly starting from the depleted region at the edge. Hence, the derivative of the electron density with respect to spatial coordinate is small, which in turn determines the width of the incompressible strip. For sufficiently weak confining electric fields, these dipole strips are so wide that they effectively isolate the compressible strips on either side, and these dipole strips earn the name incompressible strips, as they represent a strip where no screening can take place.

As mentioned, the Thomas-Fermi approximation is viable only if the potential varies slowly on the scale of magnetic length, therefore, the TFA becomes questionable if the strip width becomes comparable to $\ell_B$. The full self-consistent quantum mechanical treatment of the electron density within a mean field approximation shows that, the dipole strips cease to be incompressible for all magnetic field strengths.~\cite{siddiki2004,Suzuki93:2986} In fact, the collapse of the incompressible strips can be readily seen considering the effect of (local) external electric fields within the strip.~\cite{Sinem:11} It is straightforward to show that, the external electric field within the strip broadens the Dirac-delta shaped Landau DOS and the resulting local DOS assumes the form
\be D(E,X(E))=\frac{1}{2\pi\ell_B^2}\sum_n\frac{1}{eE_x\ell_B}|\phi_n(X(E))|^2, \label{eq:eLDOS}\ee
where $E_x=\frac{1}{e}\frac{\hbar\omega_c}{a_n}$ is the field at the strip and $X(E)$ is the normalized coordinate, that depends on energy. Hence, if the electric field at the strip becomes large (i.e. if the strip becomes narrow $a_n\sim\ell_B$, since the potential variation is fixed to $\hbar\omega_c$), then the adjacent Landau levels overlap, at the energy scales $eE_x\ell_B\sim\hbar\omega_c$. If this condition is satisfied the strip collapses, and disorder can easily scatter charge from the compressible region on one side to that on the other. In other words, the dipole strip has become too narrow and is able to screen electric fields within its width.

Such a result contrasts with 1D edge channel theories,~\cite{Buettiker86:1761} which predict more than one channel should always exist for $\nu>4$. The collapse of (narrow) incompressible strips described here is due to the finite extent of the wavefunctions. The wavefunctions can tunnel across the strip when the width of the incompressible strip becomes comparable with the quantum mechanical length scales. These scales are set by the $B$ field such as the magnetic length (width of the wavefunction) and by the Fermi wavelength, defining the thermodynamic length scale. Note that, the mentioned length scales are of the same order in typical measurements with electron densities similar to $n_{\rm el}\approx 3.0\times 10^{11}$ cm$^{-2}$ (corresponding to $\lambda_F\approx 25-35$ nm) and at intermediate $B$ fields ($B\approx5 $T, $\ell_B\approx20$ nm).

The above vanishing of the incompressible strips also applies for the CEO sample, however, one should take care of boundary conditions imposed by the sharp edge. Here, one can use the semi-classical quantization of the skipping orbits for an initial estimation of the collapse of incompressible strips.~\cite{Avishai:08,Montambaux:10} It is possible to show analytically that, the gapped regions between the levels are confined to an interval in the physical space, which is smaller than the magnetic length. Hence, electrostatically it is not possible to have incompressible strips at sharp edges. In addition to these arguments, the numerical investigations show that there are no incompressible strips at all, if an infinite wall is considered at the edges of the sample.~\cite{Wulf88:4218} We will show this situation in the following, where we also perform quantum mechanical calculations using computational techniques.

It is important to mention that, the spin generalized calculation schemes are available in the literature. In the early analytical calculations of Dempsey et al,~\cite{Dempsey:93} the effect of electron-electron interactions on spin polarization investigated, whereas Zozoulenko and co-workers utilize the density functional approach to attack the similar problem.~\cite{Igor06:075320} One can extend our above scheme to study the spin dependent behavior using these approaches. However, the important concepts of this paper do not require consideration of the spin degree of freedom, so the simpler spinless case will be addressed here.
\begin{figure}  
\centering{
\includegraphics[width=1.\linewidth]{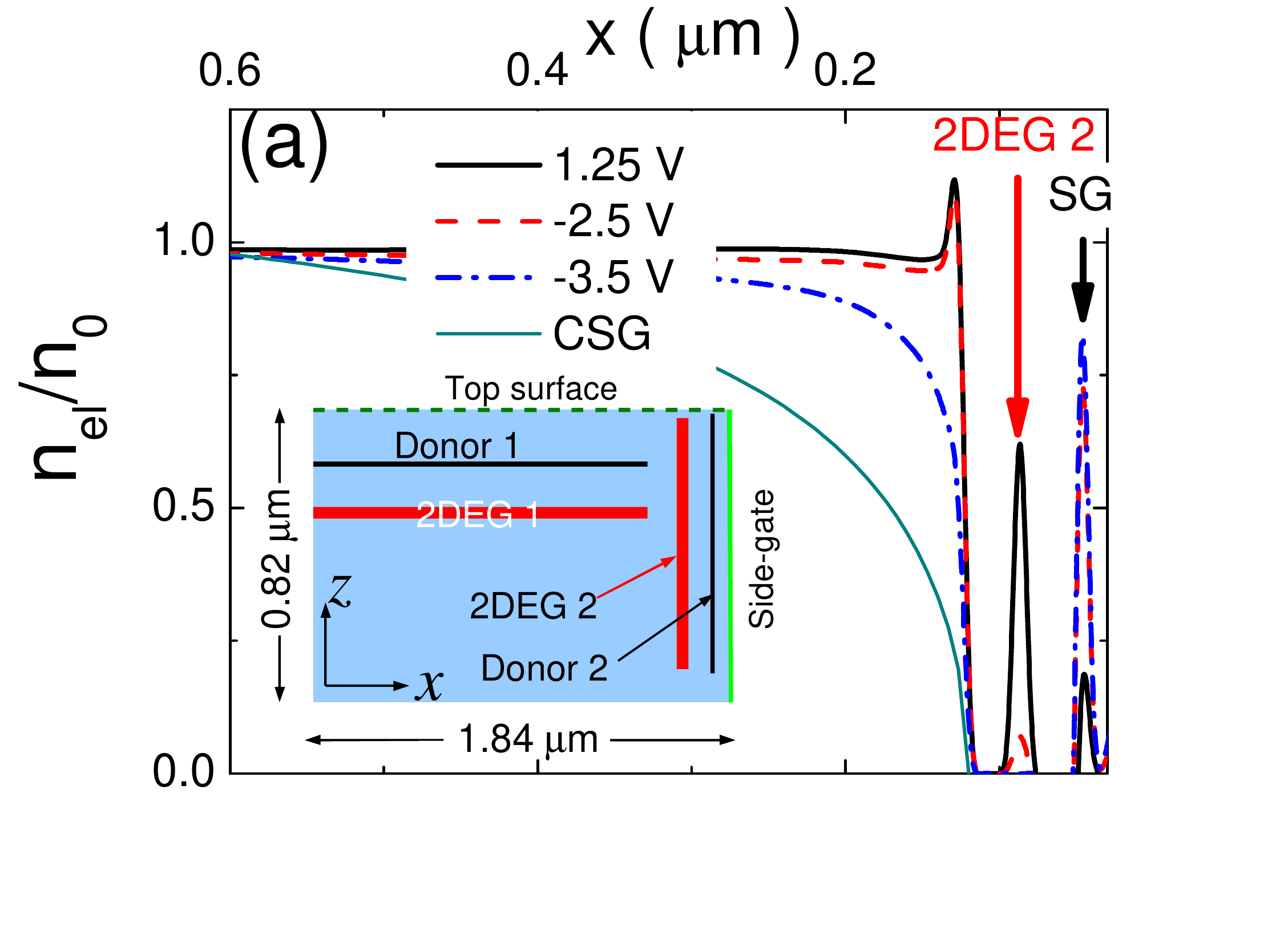}\vspace{-1.52cm}
\includegraphics[width=1.\linewidth]{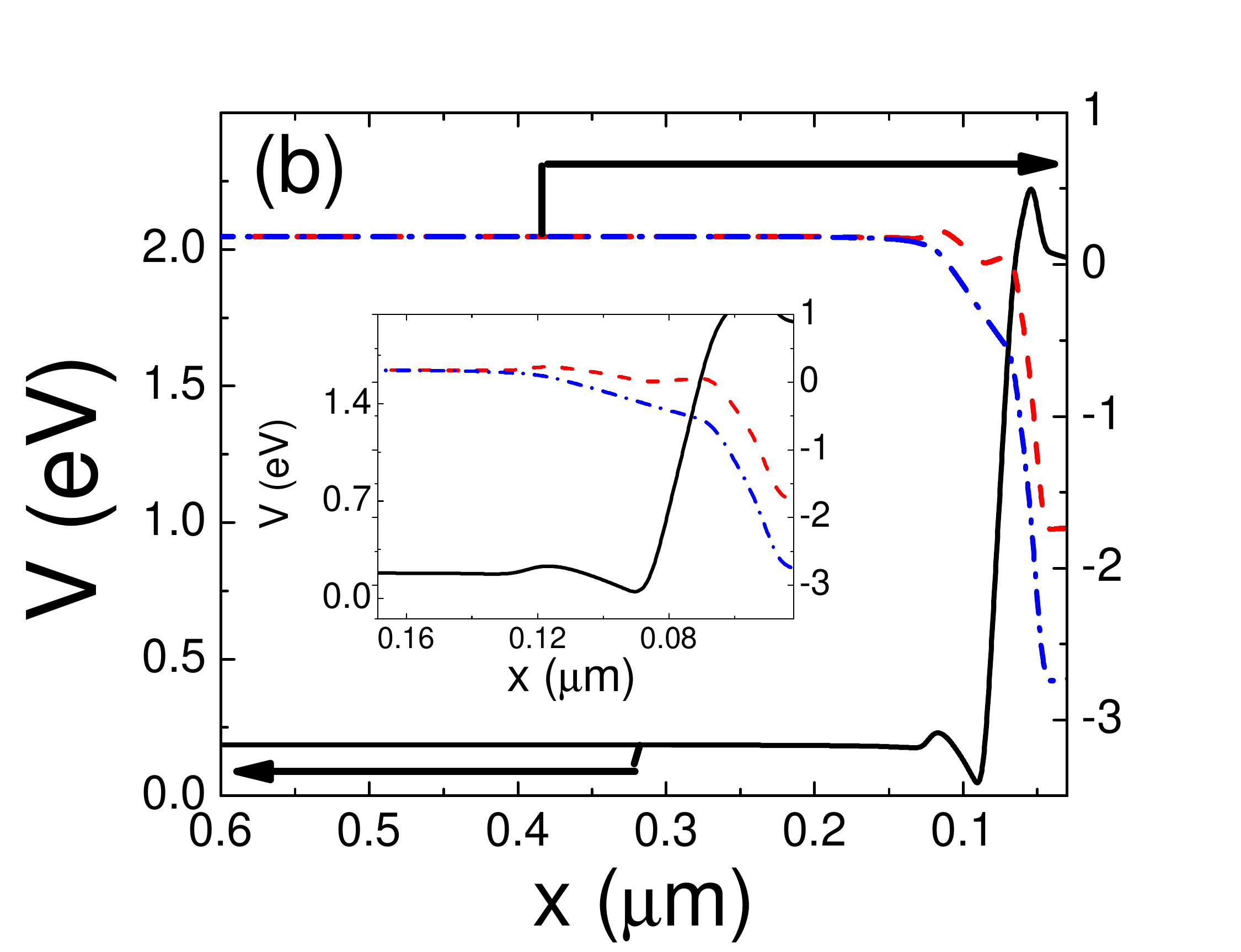}
\caption{The results of 3D calculations depicting (a) the electron density distribution of the parallel layer (2DEG 1) and (b) the total potential profile as a function of lateral coordinate. Inset in (a) shows the schematic presentation of the CEO crystal. In the first growth donors are placed 96 nm above the 2DEG, whereas in the second growth donors are place 48 nm above the second 2DEG. Both the top and side surfaces are covered by metallic gates to manipulate the electron densities, in particular the side gate allows us to change the CEO edge from sharp to smooth. The peaks at the electron density reflect the positions of the charges at the side gate (SG) and perpendicular 2DEG (2DEG 2). The inset in (b) focusses on potential distribution at the sharp edge.}\label{fig:fig6}}
\end{figure}
\subsubsection{The self-consistent numerical calculations\label{sec:NC}}
This section is devoted to investigate the effects of boundary conditions on the electronic distribution, considering the CEO sample. We solve the three dimensional Poisson equation self-consistently, in the classical domain, where finite widths of the wavefunctions are not taken into account. The boundary conditions are defined purely by electrostatics and the 2DEGs are included to the calculations as 2D charged planes. Here, we first assume vanishing temperature and zero $B$ field to obtain initial conditions. The inset of Fig.~\ref{fig:fig6}a shows a typical sample profile. We consider a crystal that is grown first in the $z$ direction (i.e. [100]), then cleaved at $x=340$ nm and the new material is grown in the $x$ direction,  [010]. Note that the side gate resides at $x=0$. The corresponding charge distribution and electrostatic potential profiles are shown in Fig.~\ref{fig:fig6}, where different potentials are applied to the side gate. The top surface of the crystal is assumed to be pinned to the mid gap of GaAs. As depicted in the schematic draw, the system is delta doped by two Silicon layers, which provide electrons for the quantum wells. The 3D Poisson equation is solved by the 4$^{th}$ order grid technique, where successive over relaxation is used for the iteration process.~\cite{Andreas03:potential,Sefa08:prb} The potential and charge distributions of the parallel electron layer presents the expected sharp variation at the cleaved edge, for small positive (solid line) and negative (broken line) side-gate potentials. At a larger negative voltage, the sharp features at the CEO edge is washed out (dash-dotted line) and the smooth edge is recovered. To compare the self-consistent density profiles with Chklovskii distribution (i.e. Eq.~\ref{eq:edensity}), we also plot the the non self-consistent density profile in Fig~\ref{fig:fig6}a (thin solid line).  One can clearly see that the Chklovskii model fails to describe the calculated edge profile even under depletion.~\cite{Oh97:13519,afif:njp2}

Once the density and potential profiles are known, it is somewhat easier to calculate the profiles at finite magnetic field. Since, the density distribution is  modified only at the incompressible regions and remains almost unchanged at compressible regions. Indeed, one can estimate the widths and the spatial positions of the incompressible strips simply from Eq.~\ref{eq:edensity}, using the electrostatic stability argument. Nevertheless, for numerical calculations one starts with $T=0$, $B=0$ profiles, increases the temperature, then adds the quantizing magnetic field via density of states and decreases the temperature step by step till the numerical accuracy is obtained. Further details of the self-consistent calculation scheme can be found in the literature, which is applied successfully to systems that assume translational invariance in $y$ direction, considering single layers~\cite{Guven03:115327,Siddiki03:125315,siddiki2004,deniz06,Afifepl:09}, double layers~\cite{siddikikraus:05,Bilayersiddiki06:,kaanPHYSEcdrag} and actual 2D systems.~\cite{SiddikiMarquardt,Sefa08:prb,Deniz10:contact} This approach, is known as the screening theory and its predictions are confirmed experimentally.~\cite{jose:epl,Mares:09,friedland,Afifcurentexp,afif:njp2,Sailer:10,josePHYSE}

The paramount outcome of the self-consistent calculations is that the 2DEG separates into compressible strips separated by dipole strips, and that only in weak confining electric fields the dipole strips are wide enough to constitute incompressible strips. The existence of the incompressible strips at low-temperatures, strong magnetic fields and high mobility samples in the small-sample limit strongly depends on the boundary conditions. Therefore, in what follows, we assume that the incompressible strip does not exist at the CEO edge, as supported by the above calculations and experiments which we discuss next. On the other hand, the properties of the incompressible strips at the ``soft" edge and bulk is determined by disorder broadened Landau levels together with temperature. In addition, our discussion on the effects of CEO edge is qualitatively independent of the properties of the soft edge, and self-consistent calculations are not necessary since the model for each edge is known from the above arguments.  Such calculations would be helpful for comparison only after a specific experimental geometry for the proposed CEO experiment is realized.

Next, we discuss the essential findings of relevant the experiments and highlight their relation to the theoretical results, mentioned above.

\begin{figure}[t]  
\includegraphics[scale=0.25]{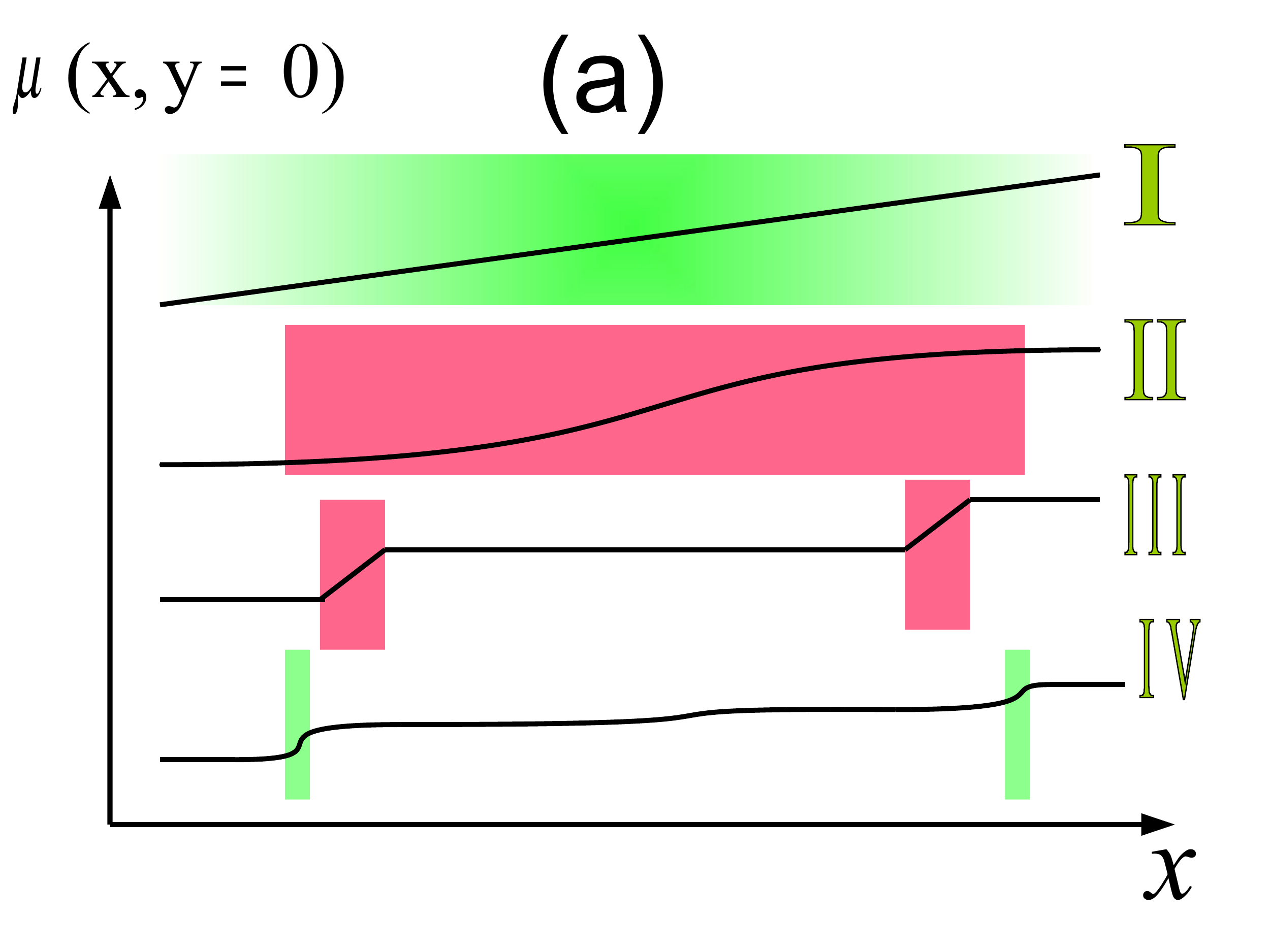}
\includegraphics[scale=0.15]{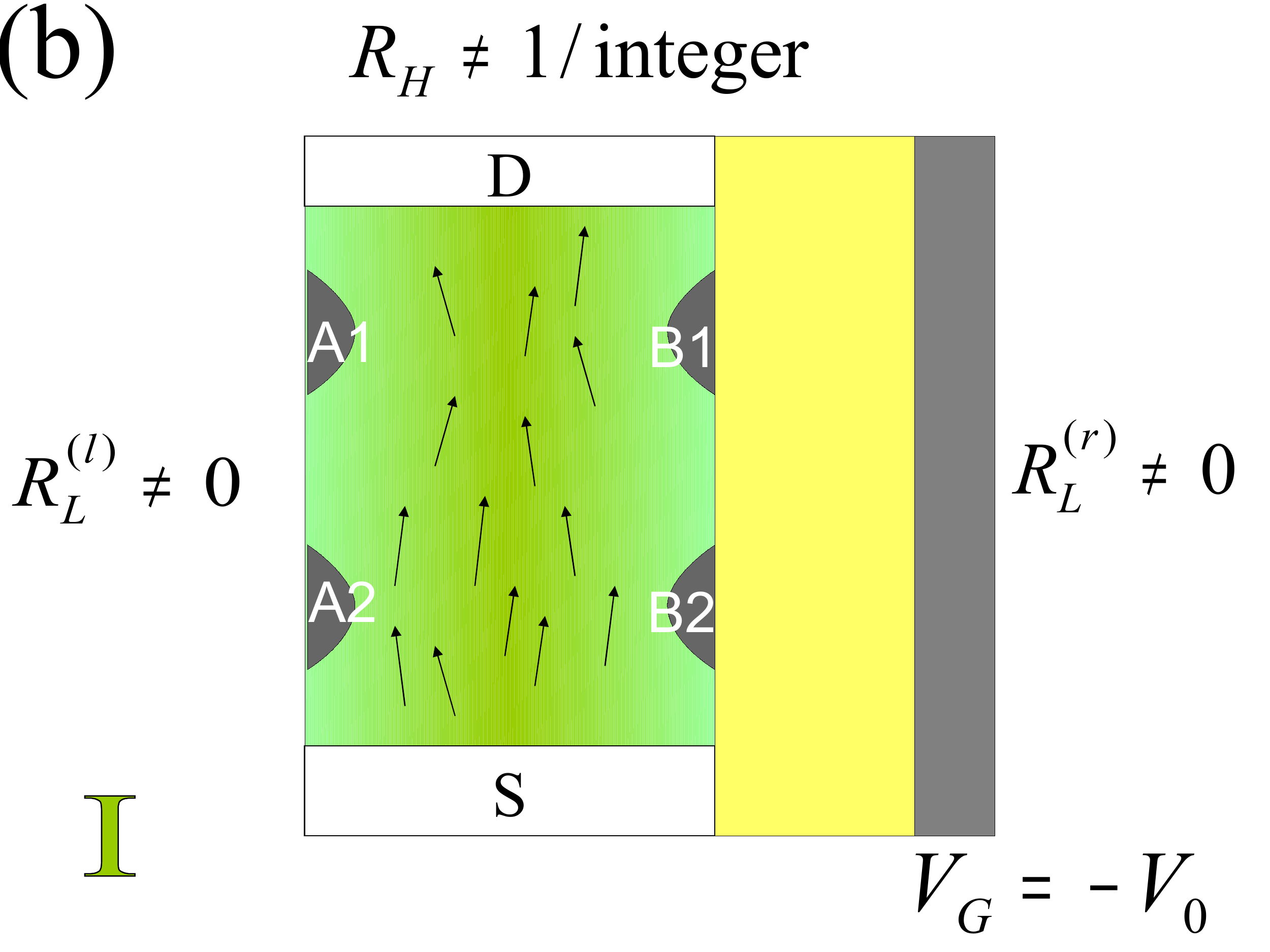}
\includegraphics[scale=0.15]{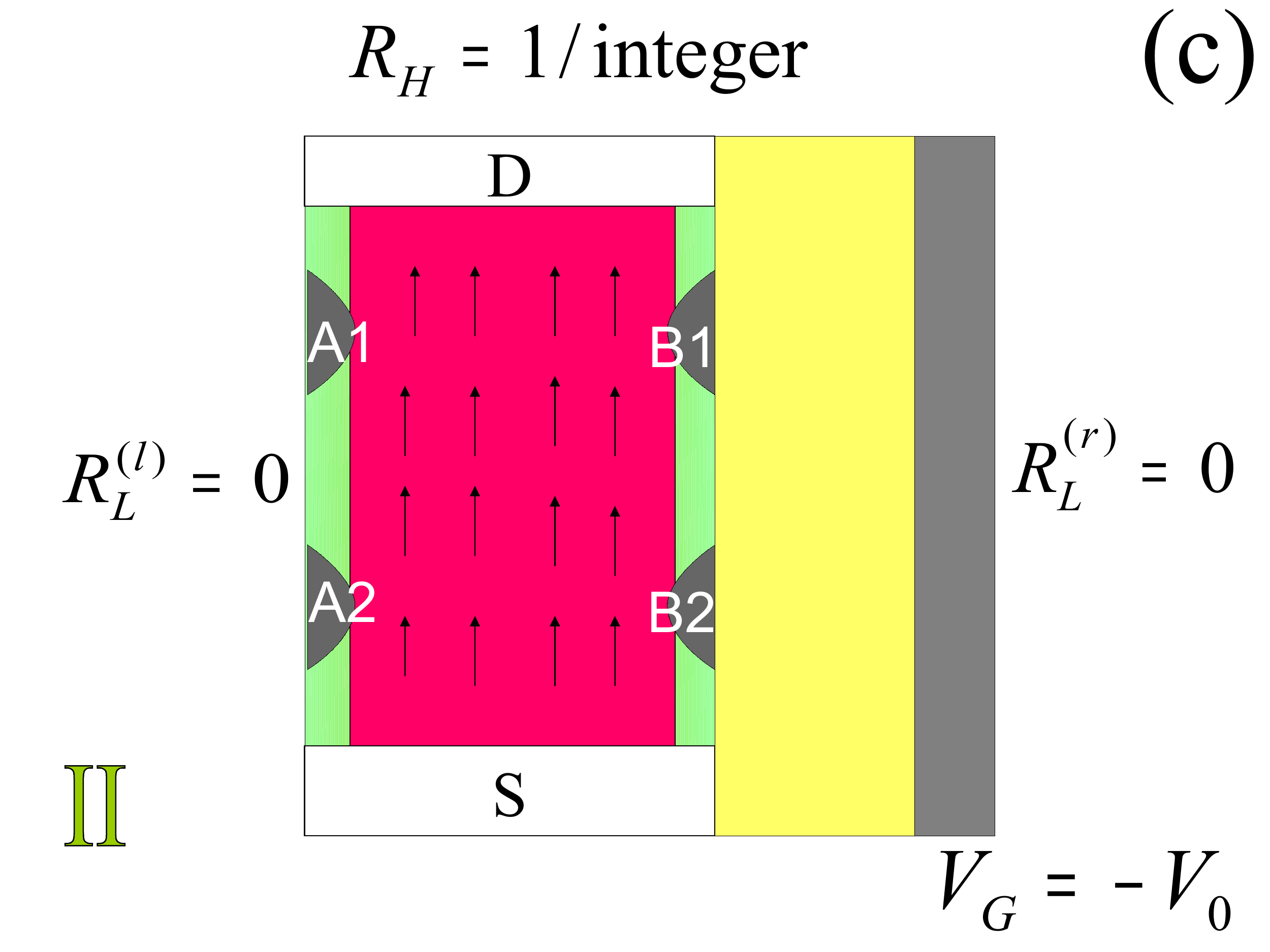}
\includegraphics[scale=0.15]{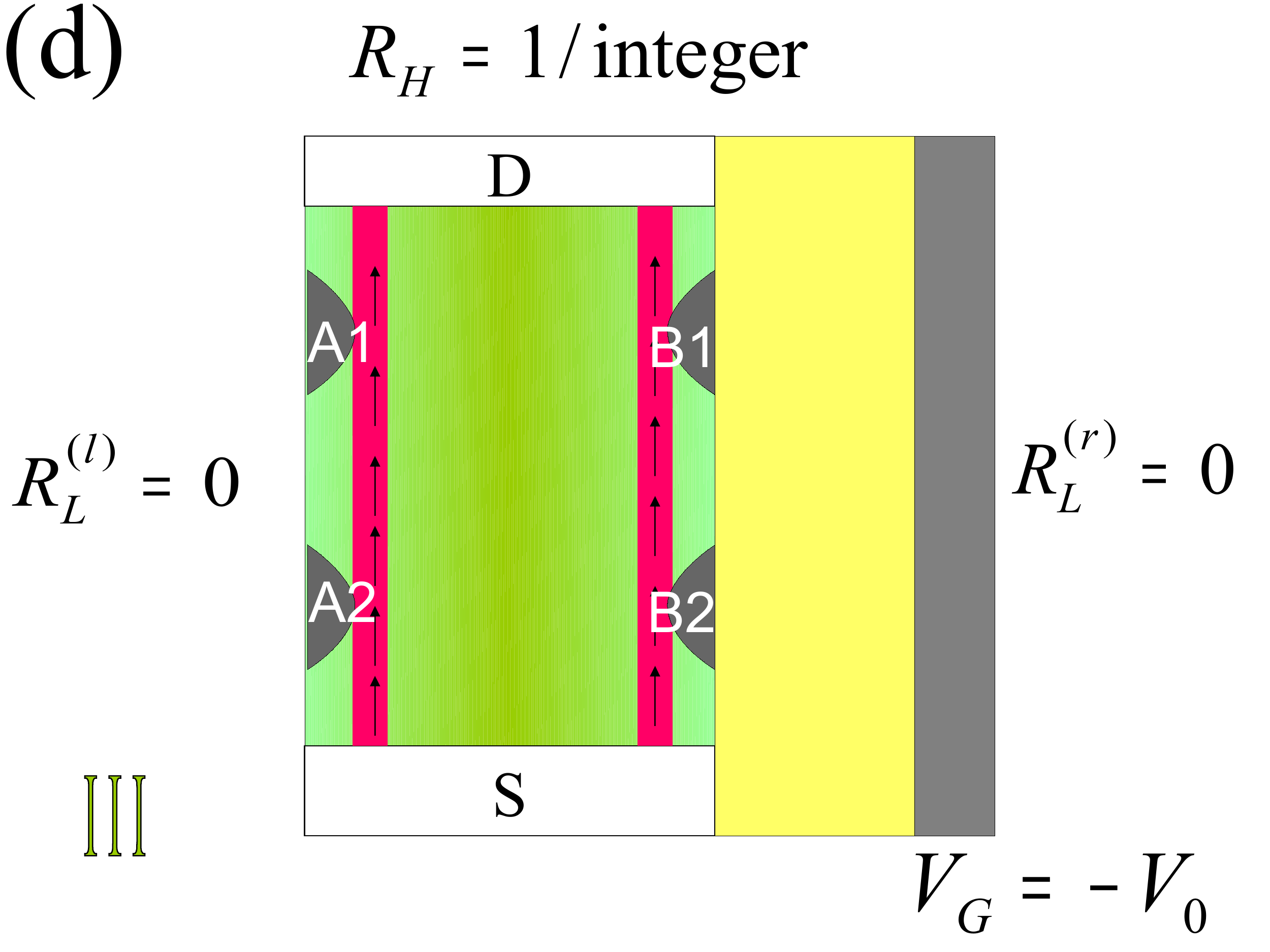}
\includegraphics[scale=0.15]{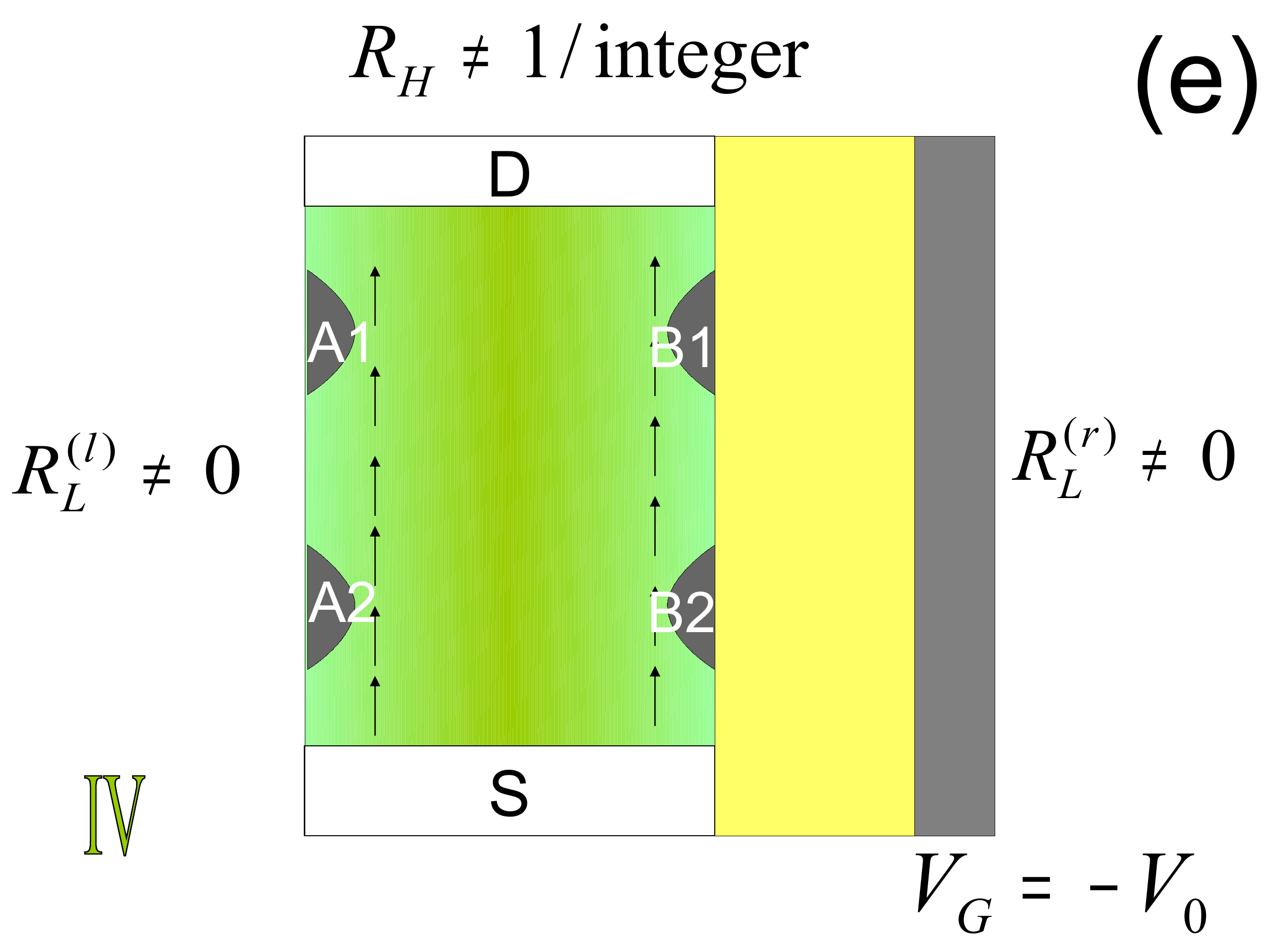}
\caption{(a) The electrochemical potential distribution at a generic sample. The smooth potential is
obtained by applying $-V_0$ to the side gate. Illustration resembles the experimental and theoretical calculations (see the related text).  The Roman numerals depict four different cases, namely: case I the system is completely compressible, case
II the bulk is incompressible, case III the incompressible
strips reside at the edges and case IV the ``evanescent'' incompressible strips reside at the edges. Shaded areas mark the current flowing regions (compressible (light-green) or incompressible (dark-pink)). The corresponding density distributions reproduced from numerical calculations, while varying the $B$ field from high to low (b-e). Color gradient codes the electron density variation at compressible regions, whereas highlighted (pink) regions are incompressible. Arrows denote the direction of excess current density. Self-consistent calculations can be found in Ref.~\onlinecite{Deniz10:contact}.}\label{fig:fig2}
\end{figure}
\begin{figure}[t]  
\includegraphics[scale=0.25]{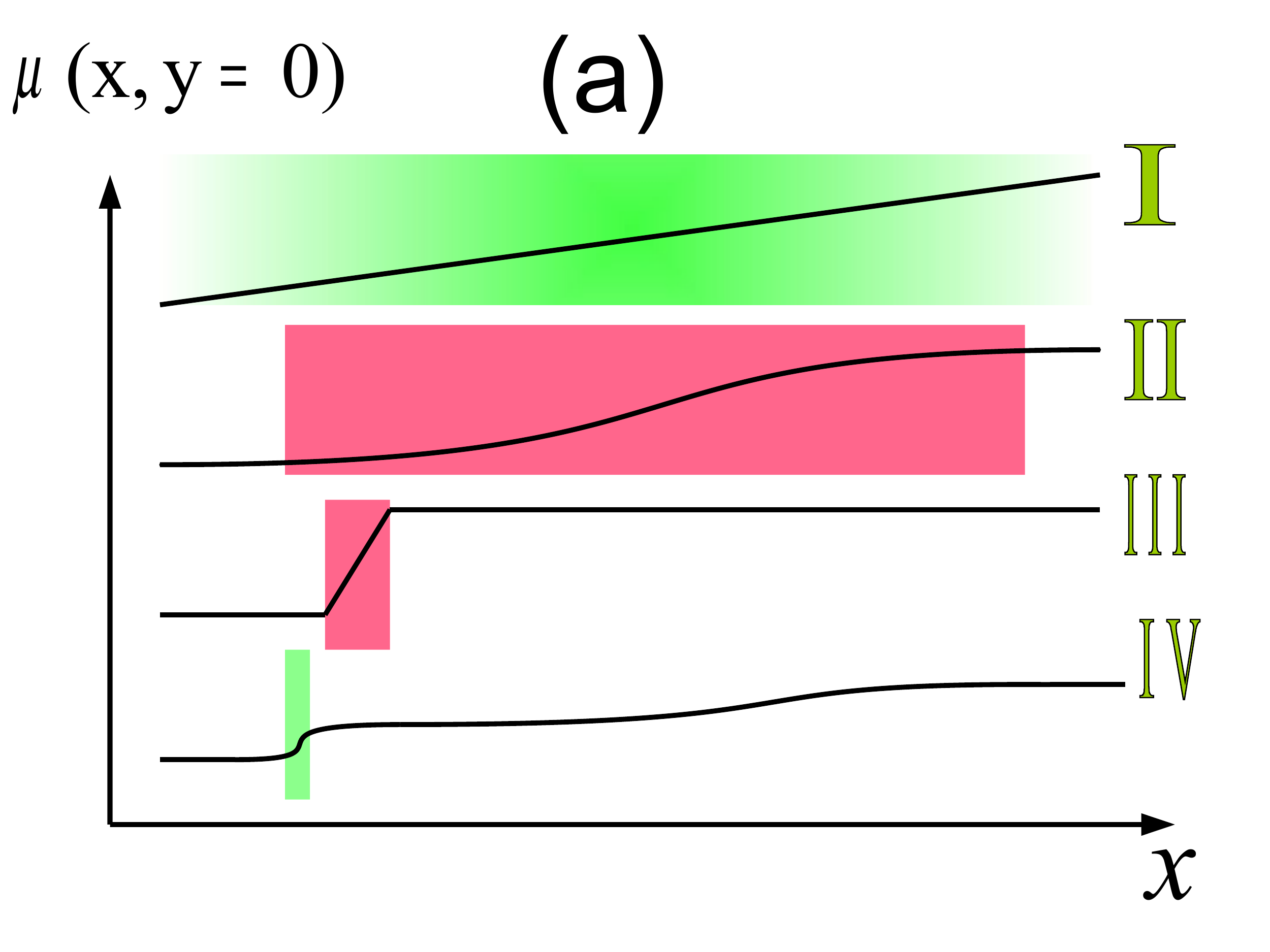}
\includegraphics[scale=0.15]{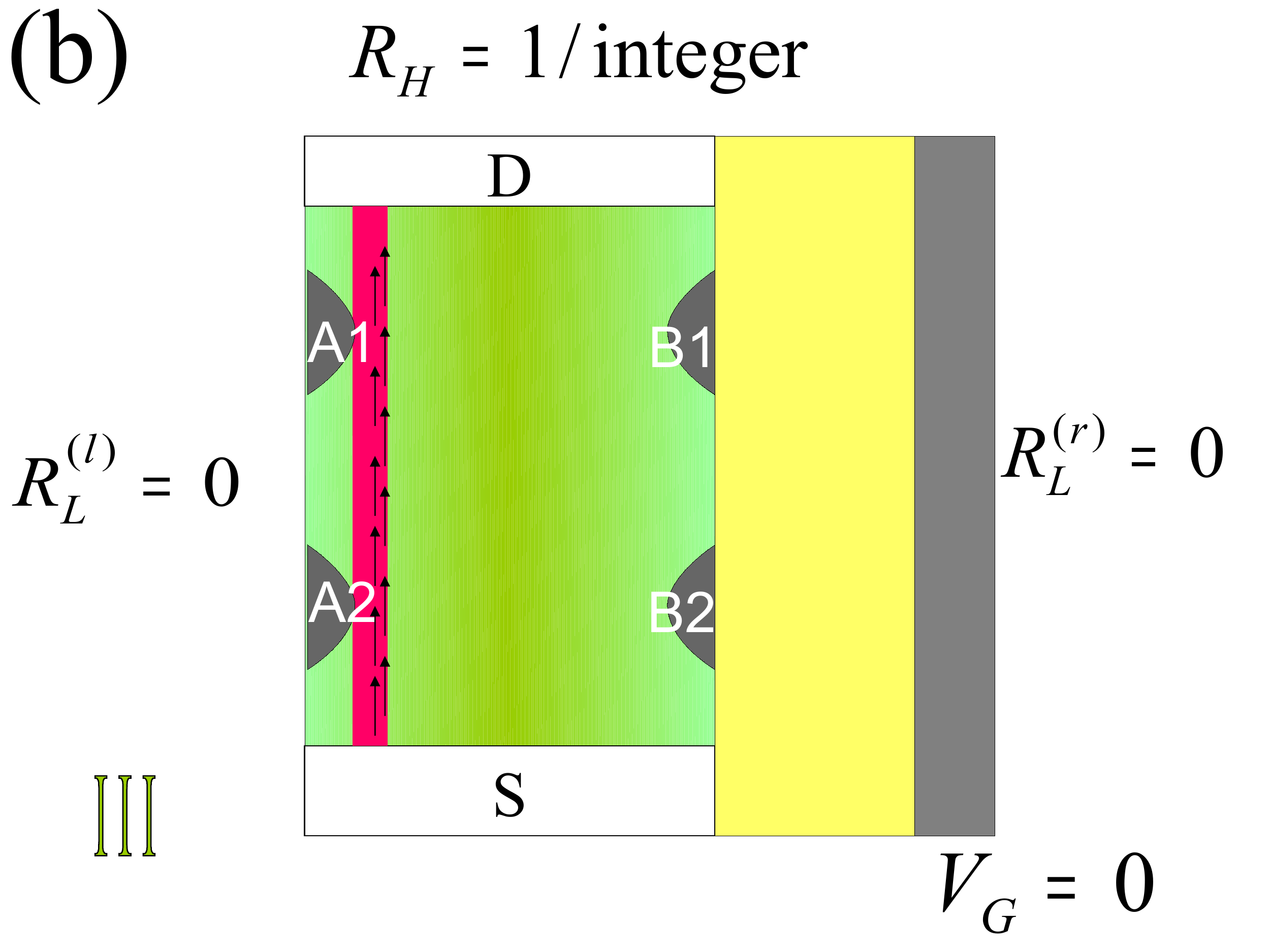}
\includegraphics[scale=0.15]{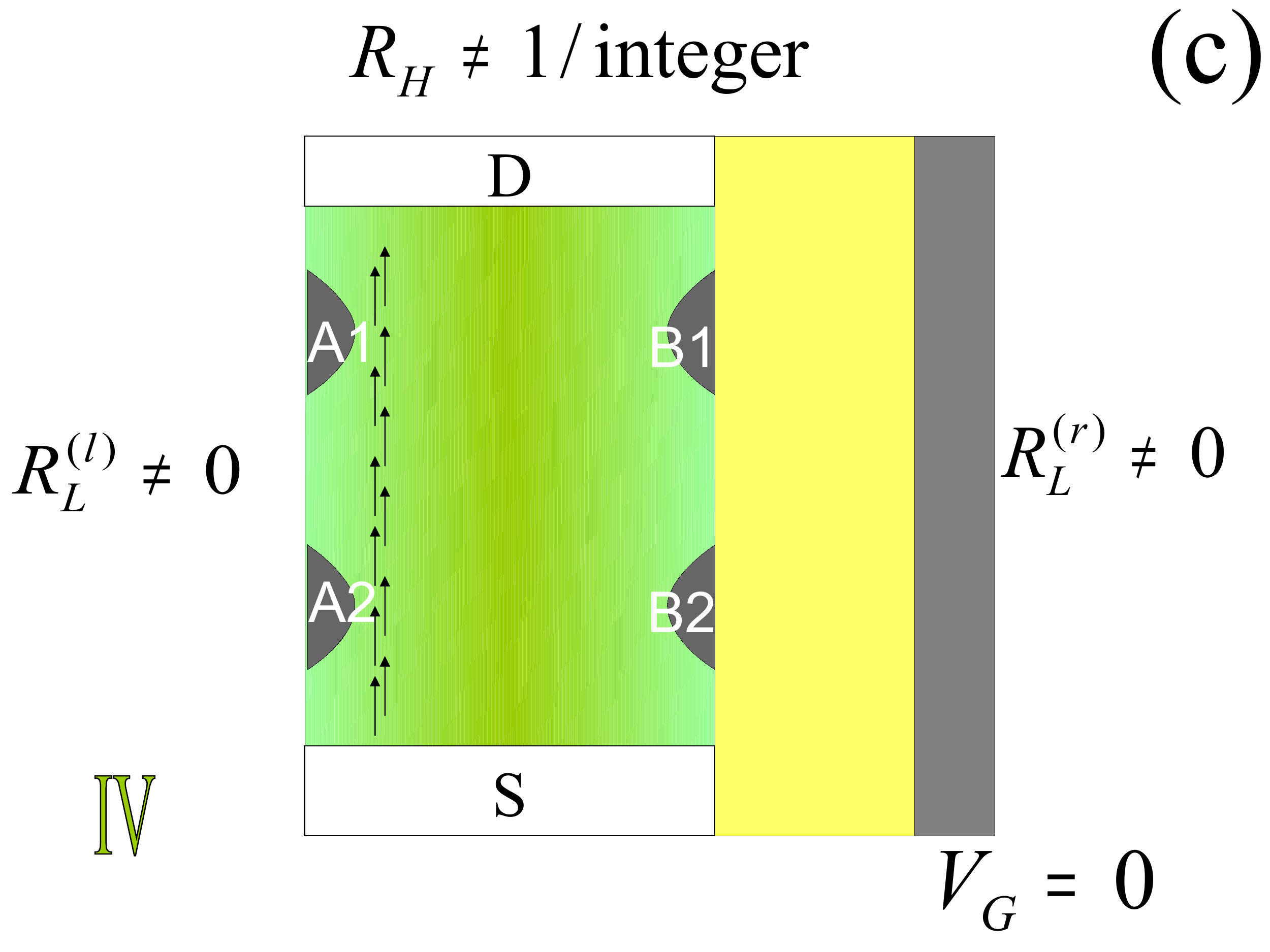}
\caption{ (a) Same as Fig.~\ref{fig:fig2}, considering a CEO sample. The spatial distribution of incompressible strips and current density for cases III and IV.}\label{fig:fig3}
\end{figure}
\subsection{Relevant experiments\label{sec:exp}}
The results of the above mentioned theoretical results are confirmed by numerous experiments, however, we would like to highlight two of them in the following. An important set of experiments is the local probe measurements of the electrochemical potential distribution considering a narrow 2DEG.~\cite{Ahlswede01:562,Ahlswede02:165,Dahlem10:contact} Here, a scanning force microscope is utilized to investigate the local potential distribution within the 2DEG. A schematic representation, resembling the experimental findings is shown in Fig.~\ref{fig:fig2}a. It is reported that, at the plateau to plateau transitions the position dependent electrochemical potential $\mu(x,y)$ varies linearly (recovering the classical Hall effect) across the sample, named as type I.~\cite{note1}. In this regime, the current flows from the entire (compressible) system as if in a metal. No incompressible regions or strips are observed. Once the bulk filling factor is in the close proximity of an integer, it is observed that, $\mu(x,y)$ varies in a highly non-linear manner, mimicking an inclined $S$. Here, the system is in the IQHE regime ($2.0\lesssim\nu\lesssim2.1$). The corresponding electrochemical potential distribution, simulating the experimental observations, is shown in Fig.~\ref{fig:fig2}a-Type II. This behavior suggests that, the current is flowing in a single bulk incompressible region.~\cite{Guven03:115327,Ahlswede02:165} We name this regime the bulk IQHE, and note that the bulk can be disorderless as shown in the figure, or it can have disorder fluctuations smaller than the gap energy. This description is consistent with the disorder picture of the IQHE~\cite{Kramer03:172}, but it is important to note that disorder is not prerequisite to observe a plateau in the quantum Hall effect. As a third type, a strong potential variation is observed at the edges, where two incompressible strips reside. Meanwhile, the potential is flat at the bulk. The observations point that the excess current is flowing from these strips. The strips are found to be parallel to the edges, far from the contacts. We call this regime to be the edge IQHE regime, where the properties of the strips are determined mainly by the electrostatics of the edges. Interestingly, one observes either a transition regime from type III to type I,  or a transition from type III to type IV and then again to type I. In the first case, the potential varies approximately in a linear manner both at the bulk and at the edges with different slopes (type IV), meanwhile global resistances show that the system is just out of the IQHE regime. In the latter case, the potential variations at the edges of type IV regime corresponds to an incompressible strip, where it is called an evanescent incompressible strip.~\cite{Sailer:10} At this interval, the strip is wider than the magnetic length ($\ell_B\lesssim20$ nm), however, narrower than the Fermi wavelength ($\lambda_F\sim25-35$ nm). The corresponding incompressible strip and current distributions are shown in Fig.~\ref{fig:fig2}b-e. The results of rigorous self-consistent calculations are published elsewhere,~\cite{Siddiki04:condmat,Deniz10:contact} however, below we will provide a brief discussion that describes how to obtain current distribution from local filling factors.

Regarding the infinite walls as boundary, the most relevant experiments are the ones that utilizes CEO edges.~\cite{Grayson05:016805,Matt:ceo1} At the CEO crystals, it is shown that no incompressible strips reside at the CEO edge, by momentum-resolved tunneling measurements. Using the perpendicular quantum well, one can tunnel into the edge of the quantum Hall bar and probe the momentum matching conditions depending on the magnetic field and the energy of the probing electrons belong to the quantum well. It is explicitly shown that, the Chklovskii incompressible strip picture~\cite{Chklovskii92:4026} breaks down at extremely sharp boundary conditions.~\cite{Matt:ceo1} The corresponding electrochemical potential distribution is shown in Fig.~\ref{fig:fig3}a where no incompressible strip resides at the CEO edge, together with the well developed incompressible strip on smooth edge (Fig.~\ref{fig:fig3}b) and evanescent strip, in Fig.~\ref{fig:fig3}c.

It is appropriate to summarize our discussion so far:\begin{itemize}\item{Dipole strips emerge as a consequence of quantizing magnetic field and direct Coulomb interactions.}\item{Their widths and existences are determined by the electrostatic boundary conditions and DOS broadening.} \item{If these dipole strips become narrower than few $\ell_B$, they cease to be incompressible and charge can scatter across them in the presence of disorder. For relatively smooth confinements, the narrow width is due to local equilibrium fields. For sharp confinements (e.g. infinite wall at a boundary), the dipole strips are never incompressible at any field range.}\item{If an incompressible region resides at the bulk of the sample and its properties are solely determined by the disorder, this situation is called the bulk IQHE. The disorder is included to our calculations via DOS broadening.}\item{If the strip is at the edge and its properties are also effected by the boundary electrostatics, this situation is called edge IQHE}\end{itemize}
\subsection{Utilizing the Ohm's law via local conductivities\label{sec:ohmslaw}}
Next, we would like to highlight the essentials of the complementary transport calculations to elucidate IQHE within the screening theory. For further details we refer to the comprehensive review by R. R. Gerhardts.~\cite{Gerhardts08:378} The self-consistent calculation of the electrostatic potential and the electron density distributions together with a local version of the Ohm's law provides an explicit relation between the formation of the incompressible strips and the quantized Hall effect.~\cite{siddiki2004} The theory elucidates all the experimental findings reported at the local probe measurements. An important aspect of the screening theory is to prescribe an explicit calculation scheme to obtain the global resistances, starting from local conductivities. Here, we briefly mention the calculation scheme to obtain local conductivities. The complementary transport calculations of screening theory assumes Gaussian single impurity potentials (i.e. disorder), as considered by T. Ando and co-workers, within the self-consistent Born approximation (SCBA).~\cite{Ando75:279,Ando74:959,Ando82:437} Since, the derivation of the DOS and conductivities has a little relevance for our discussion and is a standard procedure, we provide only the results of local longitudinal and Hall conductivities,
\begin{widetext}
\be \sigma_L(x,y)=2\frac{e^2}{h}\frac{\pi}{2}\sum_{n=0}^{\infty}\int dE[-\frac{\partial f(E,V_{\rm T}(x,y))}{\partial E}][\Gamma_n^{xx}A_n(E)]^2, \ee
and
\be \sigma_H(x,y)=\frac{e^2}{h}\nu-2\frac{e^2}{h}\frac{\pi^2}{4}\sum_{n=0}^{\infty}\int dE[-\frac{\partial f(E,V_{\rm T}(x,y))}{\partial E}]\frac{\Gamma_n^{yx}}{\hbar\omega_c}[\Gamma_n^{yx}A_n(E)]^3, \label{eq:sigmaH}\ee
where the pre-factor 2 accounts for the spin degeneracy, $n$ is the Landau index, $A_n(E)$ presents the spectral function and $\Gamma_n^{xx}$, $\Gamma_n^{yx}$ are determined by the impurity parameters. The spectral function with a broadening $\Gamma_n$ is given by,
\be A_n(E)=\frac{2}{\pi\Gamma_n}\sqrt{1-\big(\frac{E-E_n}{\Gamma_n}\big)^2}, \ee
which is centered around $\hbar\omega_c(n+1/2)$ and assumes a semi-elliptic form. Note that, each conductivity element has contributions from all levels below the Fermi energy. The above description of the conductivity tensor elements within the SCBA, is guaranteed to preserve the Onsager relations.~\cite{Ando82:437}
\end{widetext}
To simplify our discussion we assume the limit of short-range scatterers, then the coefficients assume the forms, $(\Gamma_n/\Gamma)^2=1$, $(\Gamma_n^{xx}/\Gamma)^2=n+1/2$ and $(\Gamma_n^{yx}/\Gamma)^4=n+1/2$, where $\Gamma=4n_IV_0^2/(2\pi\ell_B^2)$ is determined by the strength of the disorder via the impurity density $n_I$ and the single impurity potential amplitude $V_0$. By a straightforward substitution, the local conductivities can be prescribed in terms of the local filling factor, together with impurity parameters. Here, one can utilize the definition of the filling factor $\nu(x,y)=2\sum_{n=0}^{\infty}\int dE A_n(E) f(E-\mu(x,y))$, where $\mu(x,y)$ is the position dependent electrochemical potential. Recall that the screening theory provides local electrostatic potential and filling factors, self-consistently. Therefore, one can obtain the local conductivities within the SCBA for given system parameters. However, such a relation can also be obtained from different
approaches.~\cite{Gerhardts75:285,Guven03:115327,Tobias07:464} We use the definition of the conductivity tensor given as
\be \hat{\sigma}(x,y)=\left(%
\begin{array}{cc}
  \sigma_L(x,y) & \sigma_H(x,y) \\
  \sigma_H(x,y) & \sigma_L(x,y) \\
\end{array}\right). \ee
Since we are mainly interested in the transport properties of the incompressible strips, we confine our discussion to a situation, where the Fermi energy (or the electrochemical potential at finite temperatures and in the presence of an external current) is in the single particle gap. Then one can easily show that, $\sigma_L(\nu=\rm{integer})=0$ and $\sigma_H(\nu=\rm{integer})=\frac{e^2}{h}\nu$. This behavior can be understood by the following line of argumentation: Due to absence of available states at the Fermi energy there is no scattering within the incompressible regions, hence, the longitudinal conductivity vanishes. Meanwhile, the Hall conductivity is just proportional to the local electron density and quantum mechanical corrections vanish at integer filling factors (the second term in Eq.~\ref{eq:sigmaH}), therefore assumes a quantized value given by $\nu$. At the compressible regions, the system behaves like a metal, i.e. $\sigma_L$ is finite and $\sigma_H$ varies approximately linear with the applied $B$ field. Note that, the actual values of the conductivities at the compressible regions depend on the properties of the impurity and the approximation utilized.

Based on the above brief discussion we describe the local conductivities as \be \sigma_L(x,y)\left\{\begin{array}{cc}
  =0, & \nu(x,y)={\rm integer} \\
  \neq0, & \nu(x,y)={\rm non-integer} \\
\end{array}\right\} \nn\ee
and
\be \sigma_H=\frac{e^2}{h}\nu. \nn \ee

Given the electron density one can obtain the global resistances
using the local Ohm's law,
$\textbf{j}(x,y)=\hat{\sigma}(x,y)\textbf{E}(x,y)$, for an imposed
external current $I$. It is also important to note that, through
out this work we only deal with the excess current $I$ injected
from the contacts, however, the (chiral) equilibrium current
$I_{\rm eq}$ is not considered. We assume that the injected
current is much larger than the equilibrium current. The effect of
$I_{\rm eq}$ will be discussed elsewhere. Since the
longitudinal conductivity $\sigma_L$ vanishes once the electronic
system has a percolating incompressible strip \emph{somewhere} in the sample, simultaneously the longitudinal
resistivity and the electric field along the strip also vanishes.~\cite{Guven03:115327} Therefore, the excess current is confined to these incompressible regions, if it exists.
\begin{figure}  
\centering
\includegraphics[width=1.\linewidth]{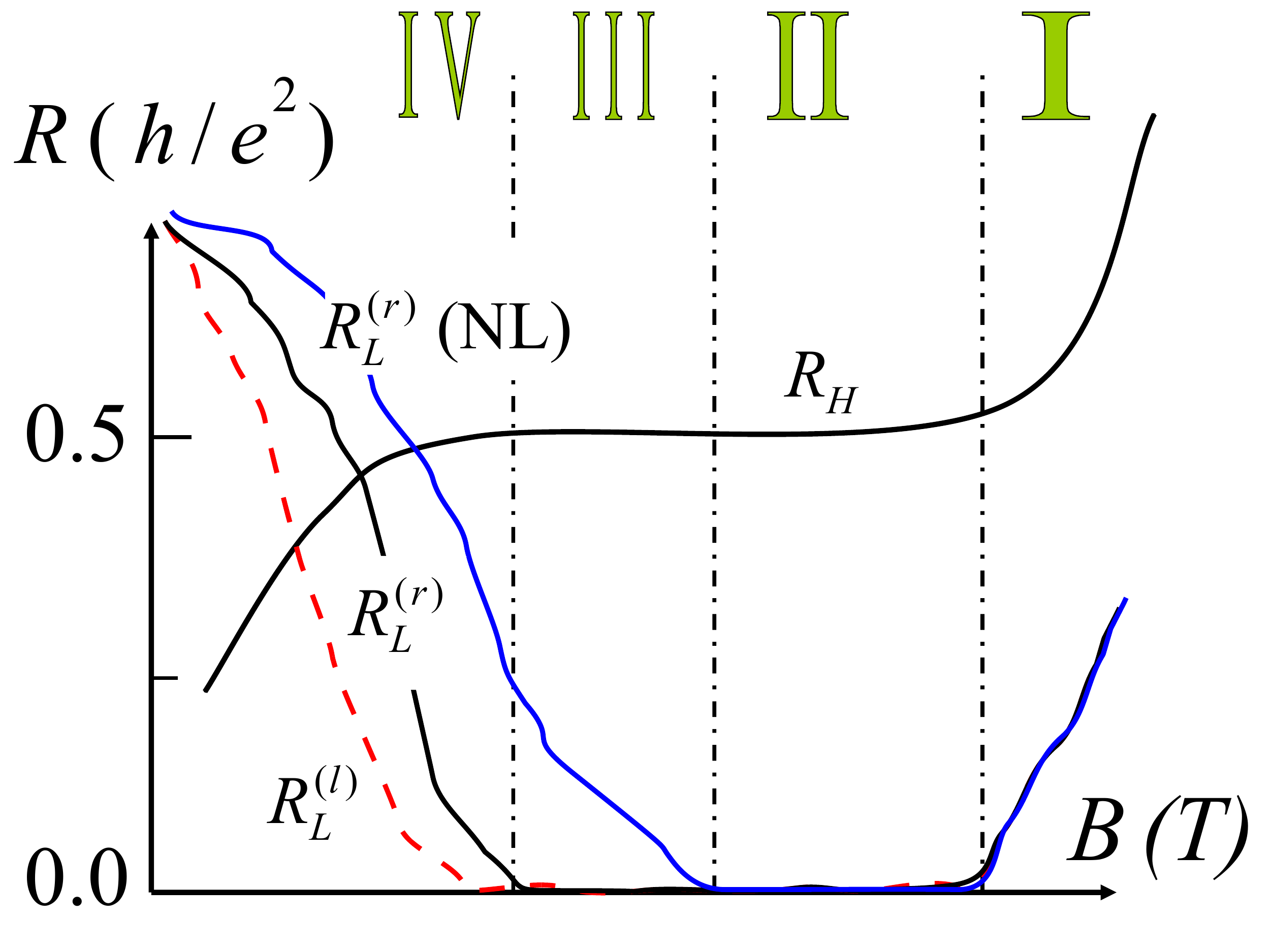}\\
\caption{The predicted global resistances as a function of magnetic
field, when the measurements are performed at sufficiently low
temperatures. The longitudinal resistances predicted at smooth edge $R_L^{(l)}$ (broken line), at sharp edge in linear transport regime $R_L^{(r)}$ (black solid) and at non-linear regime $R_L^{(r)}\rm{(NL)}$ (blue solid line) line In order to suppress the effect of long-range
potential fluctuations high mobility wafers should be
considered. Similar results obtained by self-consistent calculations at asymmetric gate defined samples can be found in Ref.~\onlinecite{SiddikiEPL:09}}\label{fig:fig4}
\end{figure}

In the following section we discuss the influence of highly asymmetric lateral confinement on the current distribution. By highly asymmetric lateral confinement we mean that, one side is CEO defined and the opposing edge is etch or gate defined. We also investigate the effects emanating from the injection contacts. We first assume that all the contacts are ``perfectly''
ideal, i.e. the edge-states are in equilibrium with the
source and drain contacts. This picture is changed
if the contacts are non-ideal with finite reflection and transmission coefficients, and then one should reconsider the
spatial distribution of the incompressible strips in front of the contacts.~\cite{Dahlem10:contact,Deniz10:contact} Therefore, one should include scattering events and/or the effect of the electric fields near
the contacts and between the boundaries of compressible and
incompressible regions.

\section{Two edge regimes at four characteristic $B$ fields, with ideal/non-ideal contacts\label{sec:contacts}}
We begin by defining ideal and non-ideal contacts before proceeding with the discussion of their influence on the current distribution at CEO samples. An ideal contact fully equilibrates all incoming incompressible strip currents with the contact such that all outgoing incompressible strips have their outermost chemical potential set by the contact chemical potential. A non-ideal contact partially reflects some of the incoming chemical potential from one or more incompressible strips to the outgoing incompressible strips. In the language of the Landauer-B\"uttiker formalism,~\cite{Buettiker86:1761} the incompressible strips are referred to as ``edge states", and the contacts are assigned reflection and transmission coefficients for each incompressible strip. These contacts thus become the defining components of the single particle non-local conductance, where dissipation takes place due to equilibration.
The next definition is that of a good ohmic contact, whose resistance is much smaller than the other measured resistances in the problem. Explicitly, a non-ideal contact can still be ohmic, if the contact resistance remains below others resistance scales. Once the contact resistance becomes dominant, then it is called a bad or non-ohmic contact. A detailed discussion on ideal/non-ideal and ohmic/non-ohmic contacts can be found in Ref.~\onlinecite{Dahlem10:contact}.

Early scanning probe experiments~\cite{Ahlswede02:165} and very recent comprehensive investigations~\cite{Goektas:Diss,Dahlem10:contact} using various experimental methods show that there is a depleted electron density just in front of the metallic contacts. This density gradient induces finite scattering between the current channel(s) and the contacts, yielding a non-ideal contact. Despite the low electron density region, it is shown that state of art contacts are perfectly ohmic.~\cite{Dahlem10:contact} It is only recently that, the self-consistent calculations by D. Eksi \emph{et al} could provide a quantitative description of the electron poor region in front of the contacts.~\cite{Deniz10:contact} This calculation scheme also presents the formation of compressible/incompressible strips in front of the contact. These non-ideal contact ideas are further supported in other recent theoretical investigations considering contacts.~\cite{Oswald:contacts,Tobias:contact}

Here we phenomenologically investigate the edge to bulk transition of the IQHE considering a sample defined on a CEO crystal, considering both ideal and non-ideal ohmic contacts. Following the discussion in Sec.~\ref{sec:exp}, we repeat the simplified definition: if the properties of the incompressible strips are defined by the edge profile, this situation is called the edge IQHE or small-sample limit. Whereas, if the incompressible strip or region resides at the bulk with conductance properties solely described by disorder, we call this situation the bulk IQHE or large-sample limit. For convenience we neglect the spin degree of freedom, since the effects we describe are universal, i.e. independent of whether Landau or Zeeman energies create the single-particle gap and subsequent
incompressible strip. Therefore we only deal with even-integer filling factors, i.e. $\nu=2k,~(k=1,2,3...)$ .
\subsection{Ideal contacts}
Once the sample geometry is given, conductance properties will be determined by the electron distribution within the sample and near the contacts. In the case of ideal contacts it is assumed that there is no density gradient in the close vicinity of the contacts. Therefore the incompressible strips are formed only due to the lateral confinement and are equilibrated with the contacts, without reflections. Hence, scattering between the current contacts (S, D) and the incompressible strips is completely suppressed. As a first step we consider such ideal contacts at the ends of our sample. We depict the positions and the existences of incompressible strips considering four characteristic $B$ field in descending order, in Fig.~\ref{fig:fig2}b-d. Here we consider a generic sample, i.e. both edges are relatively smooth. At sufficiently high magnetic fields and neglecting correlation effects, the lowest Landau level is partially occupied, Fig.~\ref{fig:fig2}b. Hence, the electronic system is completely compressible (shown by color graded regions). Consequently the global resistances $R_L$, $R_H$ are both finite and excess current is distributed all over the sample, case I. The conductivities can be described by the Drude model.~\cite{Asch:book} The small deviations of the arrow directions from the electric field direction is to demonstrate the scattering processes. Once the field is lowered, Fig.~\ref{fig:fig2}c case II, a large bulk incompressible region is formed and the excess current is confined to this region, where no (back)scattering is present. Hence, all arrows are directed along the applied current direction. The longitudinal resistances measured on both edges vanish and Hall
resistance is quantized. Lowering the $B$ field results in formation of two edge incompressible
strips (case III) and once more the imposed current is confined to these strips, Fig.~\ref{fig:fig2}d. At the lowest $B$ field considered here, the current is essentially confined to the \emph{evanescent} of the incompressible strips as depicted in Fig.~\ref{fig:fig2}e. However, some of the current is also distributed to the bulk of the sample. The ratio of current flowing from the evanescent incompressible strips and bulk decreases while lowering the $B$ field and case I is recovered before the next filling factor plateau sets in.

Next, we 
consider
the current distribution at a CEO sample considering the sharp edge on the RHS in Fig.~\ref{fig:fig3}b-c. The cases I and II remains unaffected by the sharp edge, therefore are not shown. In contrast to the smooth edge, the current distribution is strongly altered once the edge IQHE regime sets in, case III and IV. The excess current is essentially confined to the left edge. The direct measurement of such a distinction between CEO and generic samples is possible via the scanning probe experiments. Unfortunately, the experimental investigation of the electrochemical potential distribution at CEO samples is not available in the literature. Despite this fact, one can trace the signature of such an asymmetry at the potential distributions, where one edge is etched and the opposing edge is gate defined.~\cite{Ahlswede02:165,Ahlswede:Diss} One can see that, the potential distribution is symmetric if both edges are etched. Whereas, if a contact resides on one edge, the potential distribution is smoother on this side. Although, there is an evidence that the boundary conditions influence the electrochemical potential distribution, performing local probe experiments at CEO samples, will clarify our discussion. This is our first experimental proposal.

Alternately,
one can also measure the global resistances at the CEO sample, and indirectly measure the differences between the generic sample and CEO sample. We expect (almost) no differences between the global resistances when comparing the generic and the CEO sample at the highest $B$ fields, case I and II. For case I, the system is in a complete compressible state, hence, the external potential is almost perfectly screened and transport is determined by the metallic bulk. As a result, there is no apparent difference between the CEO and generic samples, when measuring the resistances. A similar argumentation also holds for case II, in contrast to the previous case in this situation the transport properties are imposed by the bulk incompressible region, which is not affected by the boundary conditions, i.e. the longitudinal resistance vanishes for both systems $R_{L}^{(g)}=R_{L}^{\rm{(CEO)}}=0$. Above arguments only hold for very high mobility samples, where no long-range potential fluctuations exists due to disorder. The situation remains unchanged for case III, since the incompressible strip on the left side of the CEO sample still decouples opposing (probe) contacts. Consider the case IV, where the edge incompressible strips at both samples become evanescent. In this case, the current distributions are effected by the edge profile. Therefore the measured resistances differ at the lower edges of the plateau. This distinction is obliterated once case I is recovered, i.e. when the evanescent incompressible strips completely vanish. Fig.~\ref{fig:fig4} shows the expected $R_H$ (thick solid line), $R_{L}^{(r)}$ (thin broken line) and $R_{L}^{(l)}$ (thin solid
line) for the CEO sample. Notice that, $R_{L}^{(l)}$ corresponds to the longitudinal resistance of a generic sample, meanwhile $R_{L}^{(r)}$ measures the transport at the CEO edge. This is an indirect way to measure the effect of different boundary conditions on transport. Again, unfortunately, it is very difficult to deposit working contacts exactly on the top of CEO edge, experimentally. Therefore, observing such a difference due to boundary conditions is obscured. As we will show below, the solution to this difficulty is hidden in the symmetry of the IQHE.

We now will discuss DC current polarity effects for distinguishing the CEO edge from the generic case.
Let us 
first
consider a generic sample, and impose a positive DC current such that the electrochemical potential looks like as it is shown in Fig.~\ref{fig:fig2}a. That is, the right side has a higher potential energy. Once the current amplitude is increased, the left incompressible strip becomes narrower, due to the fact that there are more electrons on LHS to screen the external potential compared to equilibrium. Another way to see this effect is to consider the electric field within the strip, on left side the total potential variation is $|eV_{SD}/2+\hbar\omega_c|$ and on the opposing side the variation is $|eV_{SD}/2-\hbar\omega_c|$, where we assumed that the current is shared among the two incompressible strips equally. Here, $\hbar\omega_c$ emanates from the equilibrium current. One can see from Eq.~\ref{eq:eLDOS} that, at higher the electric fields the single particle gap becomes 
effectively
smaller, 
hence, the gapped region is reduced. The 
transport
consequence 
is that
the incompressible strip
width shrinks. On the opposing edge, the incompressible strip is enlarged to 
maintain
electrostatic stability. Once the current amplitude is sufficiently large the left incompressible strip collapses. A detailed calculation can be found in Ref.\onlinecite{Sinem:11}.The effect described is a natural result of self-consistency and is predicted by numerical calculations.~\cite{Guven03:115327} If we alter the current direction, the potential distribution will look the same, however, the opposing edges will be swapped, i.e. the left side will be elevated.
Thus there is no measurable DC polarity dependence on the global resistance in generic samples.~\cite{hot-edge}

Nonetheless, 
DC polarity 
has an important influence in asymmetric samples 
such as the CEO samples.
The explicit self-consistent calculations considering an asymmetric sample predict that the
width of the
magnetic field intervals in which IQHE is observed can be tuned by changing the current direction.~\cite{SiddikiEPL:09} This prediction is tested successfully at gated samples, however, tuning the visibility of the IQHE is limited to the non-linear transport regime.~\cite{afif:njp2} Since, it is not possible to manipulate the edge steepness 
arbitrarily via gates.
In contrast to gate defined Hall bars, at CEO samples one side serves as an infinite wall, hence, one can tune the visibility of the IQHE in a wider experimental parameter window and more 
strikingly
even at the linear transport regime. From experimental point of view, it is clear that performing such experiments is highly challenging and requires expertise regarding the CEO samples. At the moment, the experimental difficulty is surpassed and preliminary results agree well with the theoretical predictions.~\cite{Matt:marchmeet}

To sum up: In the case of ideal contacts we have seen that the bulk to edge transition can be measured 
with
local probe experiments investigating the electrochemical potential profiles. However, one cannot resolve the difference between the edge (case III) and bulk (case II) regimes of the quantized Hall effect simply by measuring the global resistances. We expect to observe a difference only at the lower edge of the plateau regime. In addition, we proposed measurements of the global resistances at the CEO samples, imposing a DC current 
with opposite polarities to observe the influence of the boundary conditions on transport. To observe the different regimes of the QHE one can also consider non-ideal contacts and perform standard QHE measurements. We discuss this case in the following subsection.
\begin{figure}
\begin{minipage}[b]{0.45\linewidth} 
\centering
\includegraphics[width=4cm]{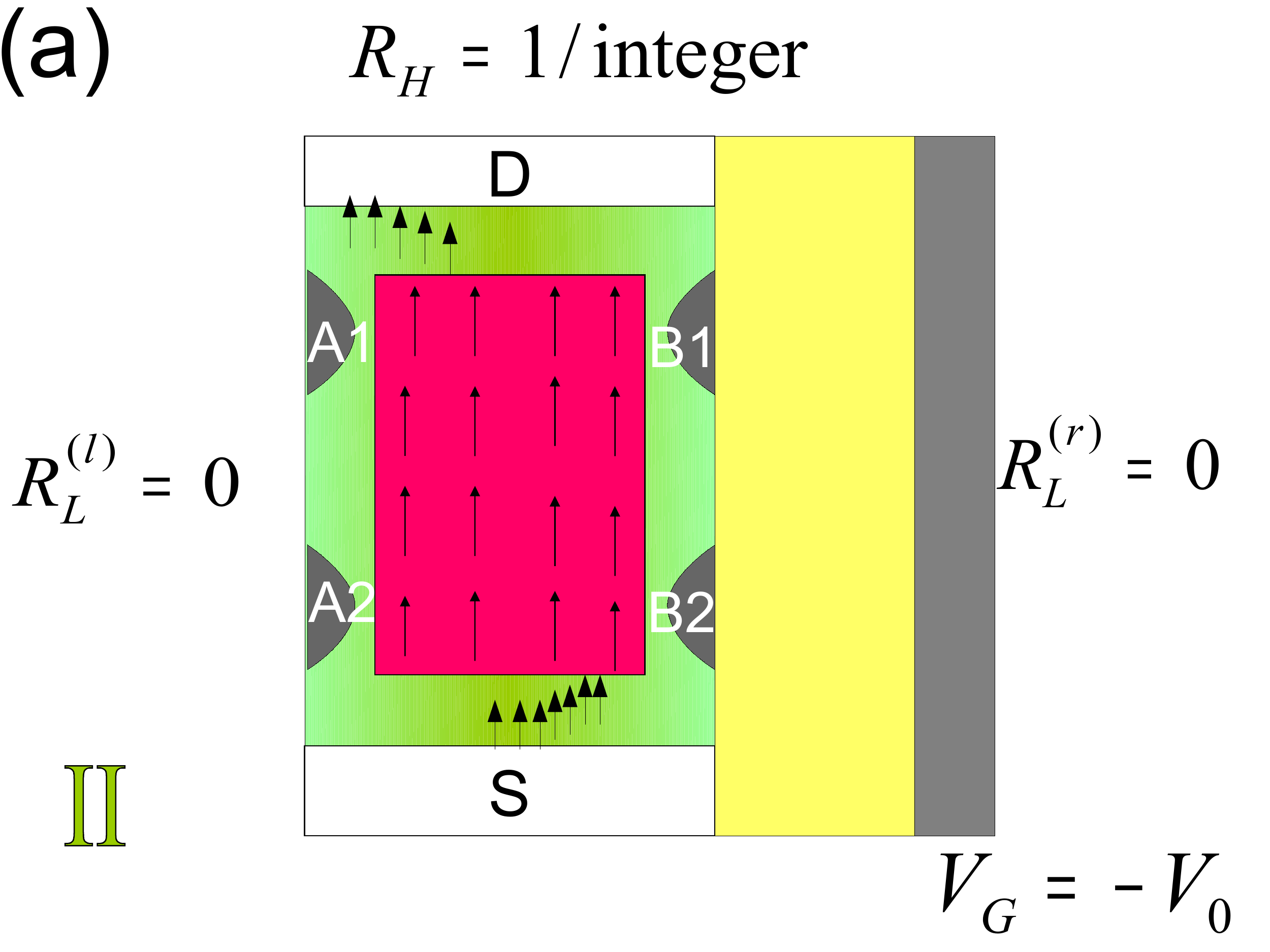}

\end{minipage}
\begin{minipage}[b]{0.45\linewidth}
\centering
\includegraphics[width=4cm]{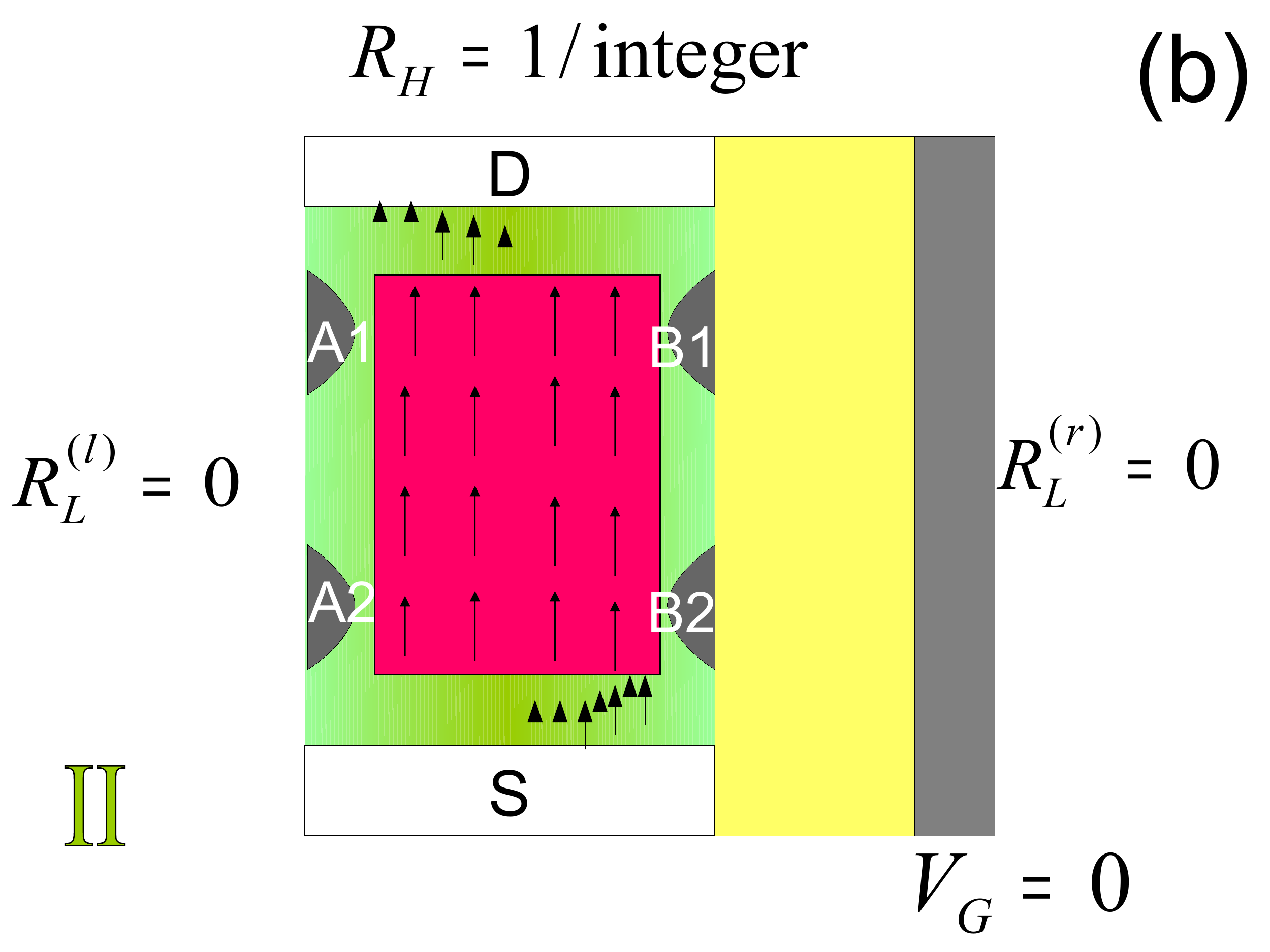}

\end{minipage}
\begin{minipage}[b]{0.45\linewidth} 
\centering
\includegraphics[width=4cm]{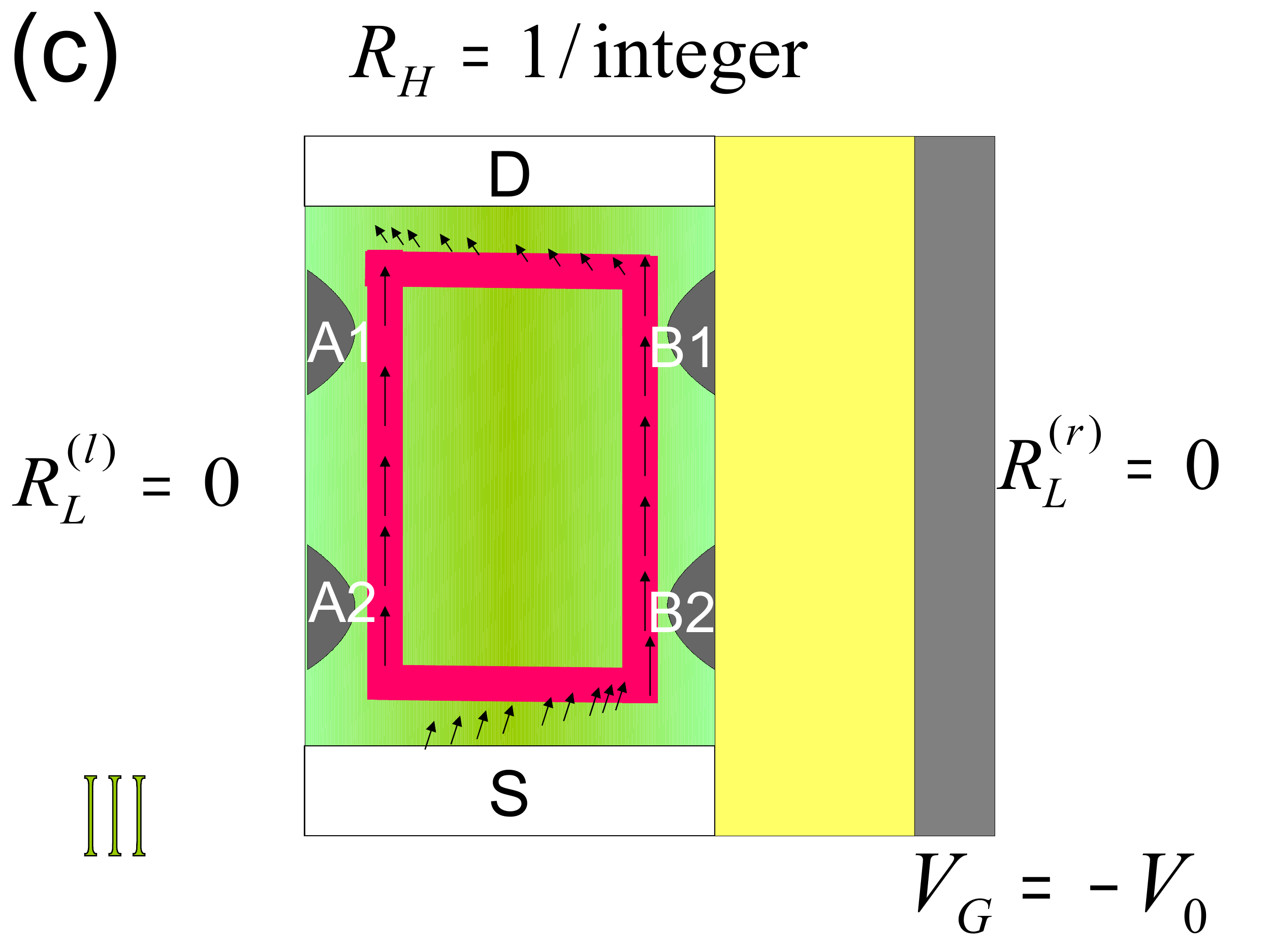}

\end{minipage}
\begin{minipage}[b]{0.45\linewidth}
\centering
\includegraphics[width=4cm]{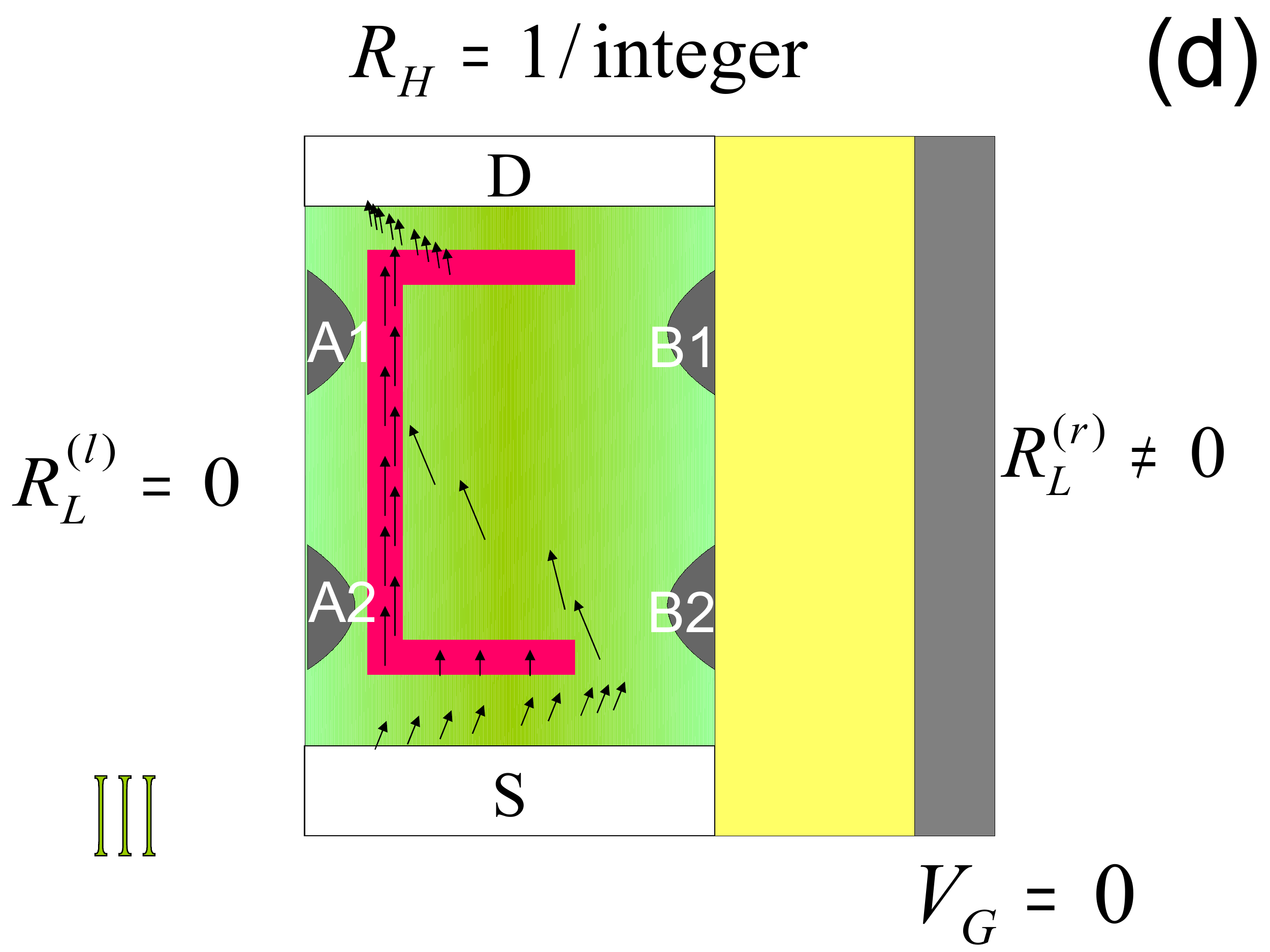}

\end{minipage}
\caption{Schematic drawing of incompressible strips and current densities considering a generic sample (left panel) and a CEO sample, while contacts are non-ideal. The location of the hot-spots depend on the field direction (see related text). The scattered arrows shown in (d) denote the bulk current, that diminishes the visibility of the quantized Hall resistance.}\label{fig:fig5}
\end{figure}
\subsection{Non-ideal contacts}
The essential difference between an ideal and non-ideal contact is the electron poor region in the close proximity of the contacts. As mentioned before, recent experimental findings show that one cannot model a contact simply by placing a metallic gate representing an equipotential on the surface. It was found that the contacts behave like fingers inserted to the 2DEG, where fingers are metallic and the space between them are insulator like.~\cite{Dahlem10:contact,Goektas:Diss} As a consequence the 2DEG is partially depleted from the contacts and a density gradient is formed in front of them. Such a density gradient leads to decoupling of the incompressible strips from the contacts. We compare the spatial distribution of the incompressible strips and current distribution for cases II and III in Fig.~\ref{fig:fig5}, also taking into account the density gradient near the contacts. The left panel depicts a generic sample, whereas the right panel shows a CEO sample. Different from the ideal contact situation, now the incompressible region (case II) or strip (case III) can only come near to the contact, whereas the region between the contact and incompressible region is compressible. This implies that scattering is finite and the transmission probability from a contact to the incompressible strip is no longer unity. Remarkably, for the non-ideal contact at the CEO sample, some of the current can also be scattered to the bulk of the sample in case III. The amount of bulk current is directly proportional to the quality of the contact. This scattered bulk current is re-collected by the incompressible strip at left, once the electrochemical equilibrium is restored. The longitudinal resistance at the left side is always zero due to the well developed incompressible strip. In contrast, one can still measure a finite $R_L$ on the right hand. Interestingly, the Hall resistance will deviate from its quantized value due to the small amount of bulk current, when measured between contacts A2 and B2. Whereas, if $R_H$ is measured between A1 and B1 it is quantized, since at this location all the current is now confined to the edge incompressible strip. Such a behavior is not observed at a generic sample 
due to lateral symmetry. In addition, if the bulk is completely incompressible (case II), then the current is directly collected by the bulk region, for both samples. This difference between case II and case III, manifests the distinction between the bulk IQHE and the edge IQHE. The former is only determined by the bulk properties of the conductivity model, namely by disorder, and the latter is determined by boundary conditions. If both edges of our sample 
are
defined by CEO, 
then one can only observe the bulk IQHE. By depositing metallic gates on both sides and varying the steepness of the edge, one 
should be able to tune the width of the IQHE plateaus by softening the sharp edges with negative gate biases.  This is our third experimental proposal.
\section{Other symmetries of the system\label{sec:other}}
In this section we consider only the non-ideal contact configuration, which is relevant for real experimental systems, and assume a clean sample without any long-range potential fluctuations %
 in the small sample limit.
 Recall that the effect of disorder is included to our model via the conductivity model we employ. The equilibration processes are investigated, based on the formation of incompressible strips. We support our arguments by the experiments that investigate the effects of incompressible strips on equilibration and decoupling of the system.~\cite{Huels:04,pan1} The term, decoupling is used to express electrochemical and thermodynamical non-equilibrium, that is the electrons cannot be redistributed easily due to the strip and have different electrochemical potentials on opposing edges.

Here, we mention two different ways of breaking symmetry of the system, 
as set by
the external $B$ field. This is different than the symmetry breaking effects due to current (electric field), discussed before. First, we investigate
hysteresis
effects that can %
arise.
We compare the decoupling properties of the edge incompressible strip with the bulk incompressible strip, hence, the bulk to edge IQHE transition. Next, using the findings of experimental and theoretical investigations considering hot-spots, we alter the direction of the external magnetic field and seek for observable differences at transport.

\subsection{Sweep direction induced hysteresis}
Let us conceive an experiment 
for the generic sample, in which one starts from low fields and ends the measurement
on an upsweep at a field such that
the lowest Landau level is partially occupied.
In such a measurement the case sequence is IV-III-II-I. First the current is flowing from both sides along the evanescent incompressible strips and the system is compressible. At a higher field (case III) the system will develop edge incompressible strips, which decouple the bulk from the edges. In this case, the bulk electrons have a different electrochemical potential compared to edges. Bulk is not accessible for the edge electrons and edge is not accessible for the bulk electrons. While increasing the $B$ field edge incompressible strips become wider to keep the electrostatic stability, until the bulk becomes completely incompressible (case II). Here, the edges are decoupled by the bulk incompressible region. Further increase of the field strength results in the disappearing of the bulk incompressible region, hence case I. If one now sweeps the field down, the opposite behavior is observed. The change of the sweep direction obviously has no influence on resistance measurements for the generic samples. Instead, one can measure the equilibration process at generic samples by means of a single electron transistor (SET) residing at the top surface.~\cite{Huels:04} It is shown that the thermodynamic equilibration is hindered by the incompressible regions and magnetic field sweep direction induces a strong hysteresis on the local electrochemical potential distribution. At temperatures below 0.1 K, the relaxation time is reported to be as long as hours. Similar observations of a strong hysteresis at the resistances of the passive layer is reported for the bilayer systems~\cite{pan1,siddikikraus:05} and is elucidated by self-consistent screening calculations.~\cite{Bilayersiddiki06:}

Now let us consider the CEO sample, where only one incompressible strip resides at the left edge. The compressible bulk is accessible for electrons to restore equilibrium, due to the absence of right strip. Therefore, while sweeping up equilibrium can be achieved easily, however, while sweeping down the bulk incompressible region hinders the equilibration. To be explicit, during down sweep the bulk incompressible region freezes the potential landscape and equilibration is suppressed. Turned around, once the bulk incompressible region is formed it stays for large $B$ intervals. At the opposite sweep direction one starts with a narrow incompressible strip at left edge, which cannot effectively decouple bulk, therefore the potential landscape is not frozen for large $B$ intervals. It is apparent that, sweep direction has an influence on the equilibration process. Consequently, down sweep presents a larger IQHE (i.e bulk and edge together), whereas up-sweep only shows the edge IQHE. This induces different paths for the longitudinal resistance, when sweeping up or down, namely a hysteresis. In addition, the visibility of this hysteresis strongly depends on the potential profile at the bulk, hence the role of long-range fluctuations has to be taken into account. Such a potential freezing model is applied to the bilayer system and the observed hysteresis is elucidated.~\cite{siddikikraus:05,Bilayersiddiki06:} There the effect of long-range fluctuations is also examined in detail, showing that the visibility of hysteresis is enhanced if the system has sufficiently strong long-range potential fluctuations. It is also discussed that, if the sample is large the edge effects are suppressed completely.
As a simple test, one can erase the memory of the system by warming up the sample during down-sweep, at the hysteresis interval. The warming process will melt the bulk incompressible region, therefore the system comes to an equilibrium state. Afterwards one can measure the $R_L$ and check if it is still in the IQHE regime. Such a test is done at the bilayer systems and it is observed that the hysteresis vanishes, similar to what we would expect for the CEO samples.

Note that, for an ideal contact one incompressible strip is sufficient to decouple opposing edges, and sweep direction is not important. However, we propose that the visibility of the hysteresis should also depend on the quality of the non-ideal contact. To be explicit, if the contact approximates to an ideal contact the hysteresis should disappear.
\subsection{The orientation of the $B$ field}

The other interesting symmetry breaking is due to the formation of the \emph{hot-spots}, known for a while experimentally~\cite{Ploog96:289,Ahlswede01:562,Dahlem10:contact} and is also calculated recently.~\cite{Deniz10:contact,Tobias:contact} The hot spot is either on the left or right bottom near the injection contact depending on the field direction. At the drain contact it is located on the diagonal corner.~\cite{note3} It is apparent that, the formation of hot-spots have no influence on the resistances measured far from the contacts at a generic sample. Remarkably, at a CEO sample the location of hot-spots alter the equilibration process of the excess current. Consider Fig.~\ref{fig:fig5}d, if $B$ field is directed along the positive $z$ axis the hot-spot is at the right bottom corner of the sample. Hence, some of the excess current can be scattered to the bulk compressible region, resulting in deviations from the quantized Hall resistance if it is measured using contacts A2-B2. Correlatively, the $R_L$ is finite on the right side. This situation is already discussed in the previous Section. Next we alter the $B$ field orientation to negative $z$ direction, then the hot-spot will form on the left-bottom corner of the sample. Consequently, the excess current is directly confined to the left incompressible strip. Hence, $R_H^{A2-B2}$ is quantized and $R_L^{B1-B2}$ vanishes. This behavior has implications on the hysteresis discussed above, since the equilibration process is also effected. Our model predicts that if a hysteresis is observed at the CEO sample, the different paths should alternate depending on the field direction, since the location of the hot-spots will also be altered and current injection process will be strongly effected.

\section{Conclusion}
In this work we investigated the effects of a sharp boundary on the transport properties and the electrochemical potential distribution of a cleaved edge overgrown sample. First, we re-introduced the calculation scheme to obtain electron and potential distributions, starting from analytical electrostatic formulation and extended our discussion to self-consistent calculation scheme. The effects of boundaries, together with interactions, on the formation of incompressible strips are discussed based on the literature and supported by self-consistent calculations. We showed that the incompressible strip does not exist at the sharp edge, agreeing with previous calculations and experiments. In the next step we summarized the essential findings of two set of experiments relevant to our discussions on the electrochemical potential distribution at narrow samples and regarding the cleaved edge overgrown samples. The basis of the complementary transport calculations of the screening theory is briefly re-introduced in Section~\ref{sec:ohmslaw}. Equipped with the theoretical and experimental findings, we investigated the current distribution comparing generic and cleaved edge overgrown samples. We considered ideal and non-ideal contacts, together with the influence of current direction on the global resistances. In Section.~\ref{sec:other}, the effects of sweep direction and $B$ field orientation is discussed. Several experimental predictions emanated through our discussions due to the sharp edge, which can be summarized as:
\begin{itemize}\item{At local probe experiments, one cannot observe a electrochemical potential drop at the sharp edge.} \item{The visibility of the IQHE should be tuned by imposing DC currents in different directions, under same experimental conditions.}\item{If both sides of the Hall bar %
are
defined by sharp edges, only the bulk IQHE can be observed.  In addition we predict that by making the edge smoother plateaus should be extended.}\item{We predict that, at the edge IQHE regime equilibration processes are promoted due to the absence of the incompressible strip at the sharp edge and are suppressed at the bulk IQHE regime by the virtue of large incompressible region at the bulk. Hence, sweeping the magnetic field direction should induce a hysteresis on global resistances, which is strongly effected by the long-range fluctuations and contact quality.}\item{Altering the orientation of the $B$ field also alters the spatial locations of the hot-spots, therefore, altering the field orientation should strongly affect the hysteresis.}\end{itemize}

Once the sharp edge behaves like a hard-wall potential, the one-dimensional edge channels proposed by Halperin~\cite{Halperin82:2185} and B\"uttiker~\cite{Buettiker86:1761} should form and the edge profile should not be important. This seems to be the case for equilibrium, namely when there is no external current. However, the screening theory states that if the width of the incompressible edge strip becomes narrower than few magnetic lengths, scattering is promoted. In the case of an external current the incompressible strip collapses, if one analytically calculates the effect of local electric fields on the local density of states. If the incompressible strip vanishes, then the longitudinal resistance becomes finite. Hence, the integer quantized Hall effect disappears.

As a final remark, recently similar asymmetries of the quantized
Hall and the longitudinal resistances were reported both
experimentally~\cite{afif:njp2,josePHYSE,jose:epl,Matt:marchmeet} and
theoretically,~\cite{SiddikiEPL:09,ozge:rect} which are all
attributed to the direct Coulomb interaction. In
Ref.~\onlinecite{SiddikiEPL:09}, it is predicted that due to the non-linear
effects induced by the imposed large current the widths of the
incompressible strips are enhanced (or reduced), which we believe
can be exploited in magnifying the asymmetrical behaviors we have
discussed.

We thank H\"useyin Kaya for his support in organizing the first ``Akyaka Nano-electronics
symposium" and the Feza-G\"ursey institute for the forth, where this work has been partially conducted. The
I.T.A.P.-Marmaris is also acknowledged for organizing the winter-school, where the discussions about utilizing the CEO crystals were initialized. This work is partially supported by T\"ubitak (109T083) and Istanbul University projects department (BAP-6970).

\begin{thebibliography}{67}
\bibitem[{not({\natexlab{a}})}]{note1}
\bibinfo{note}{For a consistency check we refer to Fig.1 of
  Ref.\onlinecite{Guven03:115327}, for the magnetic field interval $\nu<1.96$.}
\bibitem[{hot()}]{hot-edge}
\bibinfo{note}{In fact, this effect is known by the experimentalists and named
  as the hot-edge effect}.
\end{thebibliography}

\end{document}